\documentclass[twocolumn,pre]{revtex4}

\usepackage[final]{graphics}
\usepackage{amssymb}
\usepackage{amsmath}
\usepackage{latexsym}

\begin{document}

%----------------------------------------------------
%  Title
%----------------------------------------------------

\title{Test-charge theory for the electric double layer}
\author{Yoram Burak}
\email{yorambu@post.tau.ac.il}
\author{David Andelman}
\affiliation{School of Physics and Astronomy, \\
Raymond and Beverly Sackler Faculty of Exact Sciences \\
Tel Aviv University, Tel Aviv 69978, Israel}
\author{Henri Orland}
\affiliation{Service de Physique Th\'{e}orique, CE-Saclay,\\
91191 Gif-sur-Yvette cedex, France}
\date{January 19, 2004}

\begin{abstract}
We present a model for the ion distribution near
a charged surface, based on the response of the ions
to the presence of a single test particle. Near an infinite planar
surface this model produces the exact density
profile in the limits of weak and strong coupling, which
correspond to zero and infinite values of the dimensionless
coupling parameter.
At intermediate values of the coupling parameter our approach
leads to approximate density profiles that agree
qualitatively with Monte-Carlo simulation. 
For large values of the coupling parameter
our model predicts a crossover 
from exponential to algebraic decay
at large distance from the 
charged plate. Based on the test charge approach we argue
that the exact density profile is described, in this
regime, by a modified
mean field equation, which
takes into account the interaction of an ion with the 
ions close to the charged plate.
\end{abstract}

\maketitle

%===========================================================
%  Introduction
%===========================================================

\section{Introduction}

Interactions between charged objects in solution are determined 
by the distribution of ions around them. Understanding these
distributions is thus of fundamental importance for theoretical 
treatment of water soluble
macromolecules such as polyelectrolytes, charged membranes, and 
colloids \cite{LesHouches,PhysicsToday}.
In recent years, much interest has been devoted to correlation
effects in ionic solutions and to attempts to go beyond mean field
theory in their treatment. 
In particular it has been realized 
that such effects
can lead to attractive interactions between 
similarly charged objects, as was demonstrated in theoretical models 
\cite{KAJM92,StevensRobbins90,NetzOrland99,LukatskySafran99,Shklovskii,HaLiu97,
PodZeks88,Netz01},
simulation \cite{GJWL84,KAJM92,DAH03,GMBG97,MoreiraNetz02} and experiment 
\cite{KMQ88,KMSS93,Bloomfield91,Raspaud98,RauParsegian}.

Despite the theoretical interest in ion correlation effects,
they are not fully understood even in the most simple model for
a charged object surrounded by its counterions, shown schematically
in Fig.~1. The charged object in this model is an
infinite planar surface localized at the plane $z=0$,
having a uniform charge density $\sigma$.
The charged plate is immersed in a solution
of dielectric contact $\varepsilon$ and is neutralized by
a single species of mobile counterions (there is no salt in the 
solution). These counterions are confined to the
half space $z>0$ and each one of them
carries a charge $e$, which is a multiple of the unit charge for 
multivalent ions. The ions are considered as point-like, 
\textit{i.e.}, the only interactions in the system, apart from
the excluded volume at $z < 0$, are electrostatic. 

The system described above is characterized by a single dimensionless 
coupling parameter \cite{NetzOrland00} 
\footnote{A related dimensionless parameter is $\Gamma$, conventionally
defined for the two dimensional one component plasma 
(see Ref.~\cite{Shklovskii}). This parameter is related to $\Xi$ as
follows: $\Xi = 2 \Gamma^2$.
} 
\begin{equation}
\Xi = \frac{2\pi e^3\sigma}
{(\varepsilon k_B T)^2}
\label{Xidef1}
\end{equation}
where $k_B T$ is the thermal energy. At small
values of this coupling parameter the 
electrostatic interaction between ions
is weak. As a result, in the limit $\Xi \rightarrow 0$ 
mean field theory is exact, as can be formally proved
using a field-theory formulation of the problem 
\cite{NetzOrland00}. Correlations between ions close to the
charged plate play a progressively more important role with
increase of the coupling parameter. From
Eq.~(\ref{Xidef1}) one sees that this happens
with an increase of the surface charge, with 
decrease of the temperature or dielectric
constant, and with increase of the charge or, equivalently,
the valency of counterions. 
The model of Fig.~1 thus provides an elementary 
theoretical framework for
studying ion correlation effects near charged objects, with
no free parameters other than $\Xi$, which tunes and
controls the importance of ion correlations. 

\begin{figure}
\scalebox{0.45}{\includegraphics{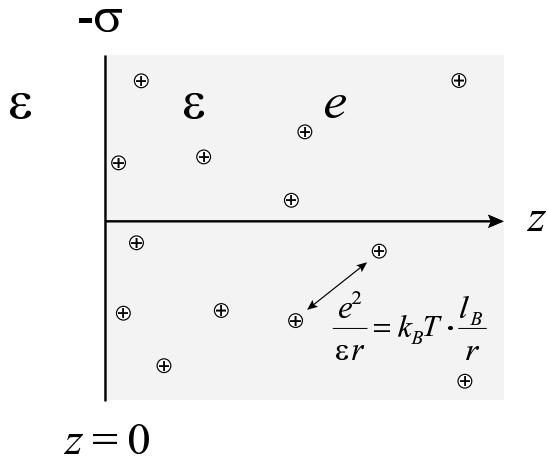}}
\caption{Schematic description of the double layer
model considered in this work. An infinite charged plate having a uniform
surface charged density $\sigma$ is immersed in a dielectric medium
having dielectric constant $\varepsilon$ on both sides of the plate.
The charge of the plate is neutralized by point-like counterions,
carrying each a charge $e$. These ions are confined to the
positive $z$ half space, where $z = 0$ is the plane occupied by 
the plate. In thermal ($k_B T$) units the interaction between two
ions is given by $l_B/r$, where $r$ is the distance between the ions
and $l_B = e^2/(\varepsilon k_B T)$ is the Bjerrum length.
}
\end{figure}
In recent years two theoretical approaches were proposed for 
treatment of the strong coupling limit $\Xi \rightarrow \infty$.
The first approach \cite{Shklovskii} is based on properties
of the strongly coupled, two dimensional 
one component plasma, and emphasizes
the possibility of Wigner crystal like ordering parallel
to the charged plane. The second approach \cite{Netz01}
is formally an exact, virial type expansion applied to 
a field-theory formulation of the partition function.
Both of these approaches predict an exponential decay
of the ion density distribution in the strong coupling
limit, leading to a more compact counterion layer
than in mean field theory. 

Although the form of the density profile is established in the two
limits $\Xi \rightarrow 0$ and $\Xi \rightarrow \infty$,
its behavior at intermediate values of the coupling parameter
is still not clear.
Liquid-state theory approaches such as the AHNC approximation 
\cite{KjelMar} can probably be used in this regime, but in practice
they were applied in the literature only to relatively small
values of the coupling parameter, usually also including ingredients 
other than those in the model of Fig.~1 -- such as finite ion size, 
added salt, or an interaction between two charged planar surfaces.
The infinite planar double layer with no added salt (Fig.~1)
was recently studied using
Monte-Carlo computer simulation \cite{MoreiraNetz02}, providing
detailed results on the counterion distribution in a wide range of
coupling parameter values. These results validate the expected
behavior in the weak and strong coupling limits. In addition they
provide new data at intermediate values of the coupling parameter,
to which theoretical approaches can be compared.

We propose, in the present work, 
a new theoretical approach for treating the distribution
of counterions near the charged plate. This approach is based on
an approximate evaluation of the response of the ionic layer,
to the presence of a single test particle. While an exact evaluation
of this response would, in principle, allow the distribution of
ions to be obtained exactly, we show that even 
its approximate calculation provides meaningful and useful results.
In the limits of small and large $\Xi$ the exact
density profile is recovered. 
At intermediate values of the
coupling parameter our approach agrees semi-quantitatively with all
the currently available simulation data.

The outline of this work is as follows. In 
Sec.~\ref{sec:model} we present
the model and discuss why it produces the exact density profile in
the weak and strong coupling limits. In  
Sec.~\ref{sec:numerical} we present
numerical results for the density profile close to the charged plate,
and compare them with simulation results of Ref.~\cite{MoreiraNetz02}.
Numerical results of our model, 
further away from the charged plate, where there
is currently no data from simulation, are presented in 
Sec.~\ref{sec:far}, 
and scaling results are obtained for the behavior of
our model in this regime. Finally, in Sec.~\ref{sec:further}
we discuss the relevance of our model's predictions, at small
and large $z$,
to the exact theory. Many of the technical details and derivations
appear in the appendices at the end of this work.

%=========================================================
\section{Model}
%=========================================================
\label{sec:model}

%---------------------------------------------------------
\subsection{Scaling}
%---------------------------------------------------------

Consider the system shown in Fig.~1, with the parameters $\sigma$, $e$,
$\varepsilon$ defined in the introduction. We will first express 
the free energy using the dimensionless coupling parameter
$\Xi$. In the canonical
ensemble the partition function can be written as follows
($z_i$ is the $z$ coordinate of the $i$-th ion):
\begin{equation}
{\rm exp}(-F_N) = 
\frac{1}{N!}\int \prod_{i=1}^{N}{\rm d}^3{\bf r}_i\,{\rm exp}\left[
-\sum_{i=1}^{N}\frac{z_i}{\mu}-
\sum_{j>i}\frac{l_B}{\left|{\bf r}_i-{\bf r}_j\right|}
\right]
\label{Fnoscale}
\end{equation}
where $l_B=e^2/\varepsilon k_B T$ is the distance at which the Coulomb 
energy of two ions is equal to the thermal energy $k_B T$, and  
$\mu = e/(2\pi l_B \sigma)$ characterizes the bare interaction of an ion
with the charged plane. These quantities, the only two 
independent length scales in the problem,
are known as the Bjerrum length and Gouy-Chapman 
length, respectively. We rescale the coordinates by the Gouy-Chapman length:
\begin{equation}
\tilde{\bf r}_i = \frac{{\bf r}_i}{\mu},
\label{scaling}
\end{equation}
yielding ${\rm exp}(-F_N) = (\mu)^{3N}{\rm exp}(-\tilde{F_N})$, where
\begin{equation}
{\rm exp}(-\tilde{F_N}) = \frac{1}{N!}
\int \prod_{i=1}^{N}{\rm d}^3\tilde{\bf r}_i\,{\rm exp}\left[
-\sum_{i=1}^{N}\tilde{z}_i-\sum_{j>i}
\frac{\Xi}{\left|\tilde{\bf r}_i-\tilde{\bf r}_j\right|}
\right]
\label{Fscaled}
\end{equation}
and the ratio
\begin{equation}
\Xi = \frac{l_B}{\mu}
\label{Xidef2}
\end{equation}
is the coupling parameter that was previously defined
in Eq.~(\ref{Xidef1}).
The requirement of charge neutrality is: $N/A = \sigma/e$, 
where $A$ is the planar area. Hence
the number of ions per rescaled unit area is equal to:
\begin{equation}
\frac{N}{\tilde{A}} = \frac{1}{2\pi\Xi}
\label{Densitynoscale}
\end{equation}
where $\tilde{A}=A/\mu^2$. The local 
density of ions in the rescaled coordinates
is equal to $\tilde{\rho}({\bf r})=\mu^3\tilde{\rho}({\bf r})$. Due to
symmetry this density depends only on $\tilde{z}$ and it is convenient
to define a normalized, dimensionless, density per unit length:%
\begin{equation}
\tilde{n}(\tilde{z}) = 2\pi l_B \mu^2 \rho = 2\pi \Xi \tilde{\rho}
\label{Densityscaled}
\end{equation}
having the property:
\begin{equation}
\int_{0}^{\infty}{\rm d}\tilde{z}\,\tilde{n}(\tilde{z}) = 1
\label{normalization}
\end{equation}
as seen from Eqs.~(\ref{Densitynoscale}) 
and (\ref{Densityscaled}).
From here on we will omit the tilde symbol ($\sim$) 
from all quantities, in order
to simplify the notations. In order to express physical quantities
in the original, non-scaled units, the following substitutions can
be used:
\begin{eqnarray}
{\bf r} & \rightarrow & {\bf r}/\mu  \\
n & \rightarrow & 2 \pi l_B \mu^2 \rho
\end{eqnarray}
We will also omit the subscript $N$ from
the free energy $\tilde{F}_N$, implying that $N$ is determined by charge
neutrality. 

%--------------------------------------------------------------
\subsection{Known results in the limits of small and large $\Xi$}
%--------------------------------------------------------------

We briefly review some known properties of the ion distribution
in the limits of small and large
$\Xi$ (for a more complete discussion, see Ref.~\cite{Netz01}).
In the limit of $\Xi \rightarrow 0$ mean field theory becomes exact.
The density profile is obtained from the Poisson-Boltzmann
(PB) equation and decays algebraically, having the form 
\cite{AndelmanReview}
\begin{equation}
n_{\rm PB}(z) = \frac{1}{(z+1)^2}
\label{npb}
\end{equation}
Within the adsorbed layer ions form a three dimensional, weakly correlated 
gas: the electrostatic interaction between neighboring ions 
is small compared to the thermal energy. 
This last statement can be verified by considering
the density of ions at contact with the plane, 
$\rho_{\rm PB}(0) = 1/(2 \pi \Xi)$ (see Eqs.~(\ref{Densityscaled})
and (\ref{npb})). 
The typical distance between neighboring ions is thus of order 
$\Xi^{1/3}$. In the non-scaled units this distance is much larger
than $l_B$, which validates the statement that ions are
weakly correlated:
$\Xi^{1/3}\mu = \Xi^{-2/3}l_B \gg l_B$. Note also that this
typical distance is small compared to the
width of the adsorbed layer (Gouy-Chapman length):
$\Xi^{1/3}\mu \ll \mu$.

In the opposite, strong coupling (SC) limit of $\Xi \gg 1$,
the density profile decays exponentially, 
\begin{equation}
n_{\rm SC}(z)={\rm exp}(-z)
\label{nsc}
\end{equation}
The width of the adsorbed layer is still of order $\mu$ in the 
non-scaled units, but is now small compared to $l_B$. 
Equation~(\ref{Densitynoscale}) indicates that the average
lateral distance between ions is then of order $\Xi^{1/2}$.
This distance is large compared to the width
of the ionic layer, $\Xi^{1/2} \mu \gg \mu$. On the other hand
it is small in units of the
Bjerrum length: $\Xi^{1/2}\mu = \Xi^{-1/2} l_B \ll l_B$. The ions
form, roughly speaking, a two-dimensional sheet and are highly 
correlated within this
adsorbed layer. The typical lateral separation between ions, 
$\Xi^{1/2}$, is
an important length scale in the strong coupling limit, and will
play an important role also in our approximated model.

At sufficiently large values of $\Xi$ it has been conjectured
(but not proved)
that ions form a two-dimensional, triangular close-packed Wigner
crystal parallel to the charged plate. Based on the melting temperature
of a two dimensional, one component plasma, one can estimate that
this transition occurs at $\Xi \gtrsim 31,000$ 
\cite{Shklovskii,MoreiraNetz02}. Furthermore, 
the ion-ion correlation function is expected to display short range order
similar to that of the Wigner crystal
even far below this transition threshold. The exponential decay
of Eq.~(\ref{nsc}) was predicted, based on these notions, in
Ref.~\cite{Shklovskii}. The same result can be obtained also in a formal
virial expansion \cite{Netz01}, which is valid for large $\Xi$
but does not involve long range order parallel to the charged plate 
at any value of $\Xi$.

Finally we note two general properties of the density profile that
are valid at any value of $\Xi$. First, the normalized
contact density $n(0)$ is always equal to unity -- a consequence of
the contact theorem \cite{CarnieChan81} (see also Appendix~\ref{ap:contact}).
Second, the characteristic width of the adsorbed layer is always of
order unity in the rescaled units.
These two properties restrict the form of the density distribution
quite severely
and indeed the two profiles (\ref{npb}) and (\ref{nsc}) 
are similar to each other
close to the charged plane. Far away from the plate, however, 
they are very different from each other: at $z \gg 1$ the probability
to find an ion is exponentially small in the SC limit, 
while in the weak coupling limit it decays only
algebraically and is thus much larger. Furthermore, in the weak
coupling case, moments of the density, including
the average distance of an ion from the plate, diverge.

Although the form of the density profile is known in the limits
of small and large $\Xi$, two important issues remain open.
The first issue is the form of the density profile at intermediate
values of $\Xi$. At coupling parameter values such as
10 or 100 
perturbative expansions around the limits of small or large
$\Xi$ \cite{NetzOrland00,Netz01} are of little use, because
they tend to give meaningful results only at small values of 
their expansion parameter. Such intermediate values are
common in experimental systems with multivalent ions, 
as demonstrated in Table~1.
Second, even at very small or very 
large $\Xi$ the respective asymptotic form of $n(z)$ may
be valid within a limited range of $z$ values. In particular,
for large $\Xi$ it is natural to suppose that sufficiently 
far away from the charged plate the density profile
crosses over from SC to PB behavior. 
Indeed, far away from the plate the ion density is small,
resembling the situation near a weakly charged surface.
The main objective of this work is to investigate these issues,
both of which necessitate going beyond the formal limits 
of vanishing and infinite $\Xi$.

\begin{table}
\caption{Characteristic values of $\sigma$, $\mu$ and $\Xi$ for
several representative macromolecules. Values of $\Xi$ are shown
for two cases: monovalent counterions (1-$e$) and 
$4$-valent ones (4-$e$). 
The Gouy-Chapman length $\mu$ corresponds
to monovalent ions. The cell membrane 
charge density is estimated assuming that 10\,\% of the lipids
in the membrane are charged.
The surface charge of DNA is estimated from the linear charge density along
the DNA contour, equal to $1/1.7$\,$e$/\AA, and assuming a radius
of 10\,\AA. For Mica full dissociation of charged groups is 
assumed. In all three cases the Bjerrum length is 
taken as $l_B = 7$\,\AA, which
corresponds to water at room temperature.} 
\begin{ruledtabular}
\begin{tabular}{lllll}
& $\sigma$($e$/\AA$^2$) & $\mu$(\AA) & $\Xi$(1-$e$) & $\Xi$(4-$e$)\\
\hline
Cell membrane & 0.002 & 10 & 0.6 & 40\\
DNA & 0.01 & 2 & 3 & 200\\
Mica & 0.02 & 1 & 6 & 400
\end{tabular}
\end{ruledtabular}
\end{table}

%-------------------------------------------------------------
\subsection{Test-charge mean field model}
%-------------------------------------------------------------

Our model is based on the following observation: the normalized 
density $n(z)$ is proportional to the partition function of a 
system where a single ion is fixed at the coordinate $z$:
\begin{equation}
n(z) = \frac{1}{Z}{\rm exp}[-F(z)]
\label{fdensity}
\end{equation}
where 
\begin{eqnarray}
& & {\rm exp}[-F(z_0)] = \frac{1}{(N-1)!}
\int \prod_{i=1}^{N-1}{\rm d}^3{\bf r}_i\,\times
\nonumber \\ & &
{\rm exp}\left(
-z_0 - \sum_{i=1}^{N-1}z_i - \sum_{i=1}^{N-1}
\frac{\Xi}{\left|{\bf r}_i-z_0\hat{\bf z}\right|}
\right. \nonumber \\ & & 
\left. - \sum_{j>i}
\frac{\Xi}{\left|{\bf r}_i-{\bf r}_j\right|}
\right)
\label{Fz0}
\end{eqnarray}
and
\begin{equation}
Z = \int_0^{\infty}{\rm d}z\,{\rm exp}[-F(z)]
\label{fnorm}
\end{equation}
where the coordinate of the fixed ($N$-th) ion 
in Eq.~(\ref{Fz0}) is $z_0\hat{\bf z}$. 
Equations~(\ref{fdensity})-(\ref{fnorm}) 
are exact and can be readily formulated also in the grand
canonical ensemble. 

In the original coordinates $F(z_0)$ is the free energy
of ions in the external potential:
\begin{equation}
\frac{z}{\mu} + \frac{l_B}{\left|{\bf r}-z_0\hat{\bf z}\right|}
\end{equation} 
exerted by the charged plane and fixed ion. 
Examination of Eq.~(\ref{Fz0}) shows that in the rescaled coordinates
these are ions of charge $\sqrt{\Xi}$ in the external potential:
\begin{equation}
\frac{1}{\sqrt{\Xi}}\left[z + \frac{\Xi}{\left|{\bf r}
-z_0\hat{\bf z}\right|}\right]
\end{equation} 

Our starting point for evaluating $n(z)$ is the exact relation
expressed by Eq.(\ref{fdensity}) but we will use a mean
field approximation in order to
evaluate $F(z_0)$. 
In this approximation the free energy is expressed as an
extremum of the following functional of $\varphi$:
\begin{eqnarray}
F_{\rm PB}(z_0) & = & \frac{1}{\Xi}
\int{\rm d}^3{\bf r}\left\{
-\frac{1}{8\pi}({\bf\nabla}\varphi)^2
-\lambda\theta(z){\rm e}^{-\varphi}
\right. \nonumber \\ & + & \left.
(\varphi-{\rm log}\lambda)\left[-\frac{1}{2\pi}\delta(z)
+\Xi\delta({\bf r}-z_0\hat{\bf z})
\right]
\right\} \nonumber \\
& - & F_{\rm self}
\label{FPB}
\end{eqnarray}
where $\varphi$ is the reduced electrostatic potential,
$\theta(z)$ is the Heaviside function, and
$F_{\rm self}$ is an infinite self energy which does not depend
on $z_0$. The derivation of Eq.(\ref{FPB}) is given in 
Appendix~\ref{ap:freeenergy}.

The mean field equation for $\varphi$ is found from the requirement 
$\delta F_{\rm PB}/\delta \varphi({\bf r}) = 0$:
\begin{equation}
-\frac{1}{4\pi}{\bf\nabla}^2\varphi = 
\lambda\theta(z){\rm e}^{-\varphi}
-\frac{1}{2\pi}\delta(z) + \Xi
\delta({\bf r}-z_0\hat{\bf z})
\label{PBeq}
\end{equation}
This equation describes the mean field distribution of ions 
in the presence of a charged plane of uniform charge density 
$-1/(2\pi)$ (second term in Eq.~(\ref{PBeq})) and a point charge
$\Xi$ located at ${\bf r} = z_0\hat{\bf z}$ (third term in
Eq.~(\ref{PBeq})). In cylindrical coordinates the solution 
$\varphi$ can be written as a function only of the radial
coordinate $r$ and of $z$, due to the symmetry of rotation around the
$z$ axis.

It is easy to show that at the extremum of $F_{\rm PB}$ the overall 
charge of the system, including the charged surface, test charge
and mobile counterions, is zero. 
The fugacity $\lambda$ has no effect on the extremal 
value of $F_{\rm PB}$; changing its value only shifts
$\varphi({\bf r})$ by a constant.

Equations (\ref{fdensity}) and (\ref{fnorm}), 
together with the mean-field approximation for $F(z_0)$ 
given by Eqs.~(\ref{FPB}) and (\ref{PBeq}) constitute the approximation
used in this work:
\begin{equation}
n(z) = \frac{1}{Z}{\rm exp}[-F_{\rm PB}(z)]
\label{model}
\end{equation}
\noindent where
\begin{equation}
Z = \int_0^{\infty}{\rm d}z\,{\rm exp}[-F_{\rm PB}(z)]
\label{modelnorm}
\end{equation}
We will refer to this approximation as the test-charge mean field (TCMF) 
model.

%----------------------------------------------------------
\subsection{Limits of small and large $\Xi$}
%----------------------------------------------------------
%
\begin{figure}
\scalebox{0.45}{\includegraphics{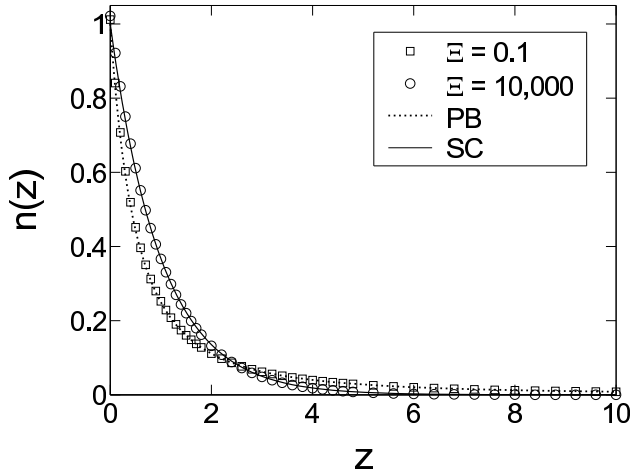}}
\caption{Density profiles, $n(z)$, numerically 
calculated using the TCMF model of 
Eqs.~(\ref{PBeq})-(\ref{modelnorm}), with $\Xi=0.1$ (squares)
and $\Xi = $10,000 (circles). The solid lines show the exact asymptotic
profiles in the low coupling, 
$n_{\rm PB}(z) = 1/(z+1)^2$, and in the strong coupling limit,
$n_{\rm SC} = {\rm exp}(-z)$.
}
\end{figure}
As a first example we present, in Fig.~2, density profiles obtained numerically
from Eqs.~(\ref{model})-(\ref{modelnorm}) at $\Xi = 0.1$ (circles) 
and at $\Xi = 10000$ (squares). The continuous lines are the 
theoretically predicted profiles at $\Xi \rightarrow 0$,
$n_{\rm PB}(z)=1/(z+1)^2$, 
and at $\Xi \rightarrow \infty$, $n_{\rm SC}(z)={\rm exp}(-z)$.
The figure demonstrates that the 
weak coupling and strong coupling limits are reproduced correctly in
our approximation. Before presenting further
numerical results, we discuss the behavior of our model 
in the two limits of small and large $\Xi$.

Our discussion is based on the following exact identity:
\begin{equation}
\frac{{\rm d}}{{\rm d}z_0}{\rm log}\left[n(z_0)\right] = 
-\frac{\rm d}{{\rm d}z_0}F(z_0) = 
-\left.\frac{\partial}{\partial z}
\left<\varphi({\bf r}; z_0)\right>
\right|_{\displaystyle {\bf r}=z_0\hat{\bf z}}
\label{dlnrhoex}
\end{equation}
where $\left\langle\varphi({\bf r}; z_0)\right\rangle$ is the thermally 
averaged electrostatic potential at ${\bf r}$, when a test
charge is \textit{fixed} at $z_0\hat{\bf z}$ (the first argument 
of $\left\langle\varphi({\bf r}; z_0)\right\rangle$
designates the position ${\bf r}$ where the potential is evaluated, 
while the second argument designates the position of the test charge,
$z_0 \hat{\bf z}$). In other words, the
gradient of ${\rm log}[n(z_0)]$ is equal to the 
electrostatic force acting on a test charge positioned at
${\bf r}=z_0\hat{\bf z}$.
This equation does not involve any approximations and is
proved in Appendix~\ref{ap:derivation}. 

Within our approximation, where $F(z_0)$ 
is replaced by $F_{\rm PB}(z_0)$, a similar 
equation holds (also proved in the Appendix):
\begin{equation}
\frac{{\rm d}}{{\rm d}z_0}{\rm log}\left[n(z_0)\right] = 
-\frac{\rm d}{{\rm d}z_0}F_{\rm PB}(z_0) = 
-\left.\frac{\partial}{\partial z}\varphi({\bf r}; z_0)
\right|_{\displaystyle {\bf r}=z_0\hat{\bf z}}
\label{dlnrho}
\end{equation}
where $\varphi({\bf r}; z_0)$ is now the solution of Eq.~(\ref{PBeq}). 
In other words, the gradient of 
${\rm log}[n(z_0)]$ is equal to the electrostatic
force experienced by a test charge positioned at ${\bf r}=z_0\hat{\bf z}$,
evaluated using the mean field equation (\ref{PBeq}). 
This quantity,
\begin{equation}
f(z_0) \equiv \left.\frac{\partial}{\partial z}\varphi({\bf r}; z_0)
\right|_{\displaystyle {\bf r}=z_0\hat{\bf z}}
\label{fofz}
\end{equation}
will be studied in detail below because of its important role within our
model. With this notation the relation between $f(z)$ and $n(z)$ reads:
\begin{equation}
\frac{{\rm d}}{{\rm d}z}{\rm log}\left[n(z)\right] = 
- f(z)
\label{dlnrho1}
\end{equation}

Using Eq.~(\ref{dlnrho1}) we can understand why both the weak and strong 
coupling limits are reproduced correctly in our model:

\subsubsection*{Weak coupling}
In the limit $\Xi\rightarrow 0$,
\begin{equation}
\frac{\partial}{\partial z}\varphi({\bf r}; z_0) 
\rightarrow \frac{\rm d}{{\rm d}z}\varphi_{\rm PB}(z).
\label{weaklimit}
\end{equation}
where $\varphi_{\rm PB}(z)$ is the solution of Eq.~(\ref{PBeq}) 
without a test charge, \textit{i.e.}, setting $\Xi = 0$.
We note that the potential $\varphi$ (Eq.~\ref{PBeq}) has three sources:
the charge of mobile counterions, $\lambda\theta(z){\rm e}^{-\varphi}$,
the uniformly charged plane, and the test charge. Although the potential
due to the test charge is infinite at ${\bf r}=z_0\hat{\bf z}$, its
derivative with respect to $z$ is zero and has no contribution in
Eq.~(\ref{weaklimit}). Using 
Eq.~(\ref{dlnrho1}) we find:
\begin{equation}
\frac{{\rm d}}{{\rm d}z}{\rm log}\left[n(z)\right] = 
-\frac{{\rm d}}{{\rm d}z}\varphi_{\rm PB}(z)
\end{equation}
This equation, together with the normalization requirement for $n(z)$
leads to the result:
\begin{equation}
n(z) = \frac{1}{Z_0}{\rm exp}[-\varphi_{\rm PB}(z)] = 
n_{\rm PB}(z)
\end{equation}

\subsubsection*{Strong coupling}

In the strong coupling limit, $\Xi\rightarrow\infty$, a correlation hole 
forms in the distribution of 
mobile counterions around the test charge at
${\bf r} = z_0\hat{\bf z}$. The structure and size of this hole,
as obtained from Eq.~(\ref{PBeq}), will be
discussed in detail later. For now it is sufficient to note 
that the correlation
hole gets bigger with increasing $\Xi$. As $\Xi\rightarrow\infty$ 
the force at $z_0\hat{\bf z}$ due to the mobile counterions 
vanishes, leaving only the contribution of the
charged plane:
$
(\partial/\partial z)\varphi({\bf r}; z_0,\Xi) 
\rightarrow 1
$.
Hence in this limit
\begin{equation}
\frac{{\rm d}}{{\rm d}z}{\rm log}\left[n(z)\right] = -1
\end{equation}
leading to the strong coupling result:
\begin{equation}
n(z) = {\rm exp}(-z)
\end{equation}
where the prefactor of the exponent follows from the normalization
condition, Eq.~(\ref{normalization}).

In the rest of this work we will explore predictions
of the TCMF model at intermediate coupling, where
neither of the two limits presented above is valid.
Before proceeding we note that a similar discussion
as above, of the weak and strong coupling limits, applies also
to the exact theory, because of Eq.~(\ref{dlnrhoex}). 

%=============================================================
\section{Numerical Results and Comparison with Simulation}
%=============================================================
\label{sec:numerical}
\subsubsection*{Results for $f(z)$}
\begin{figure*}
\scalebox{0.45}{\includegraphics{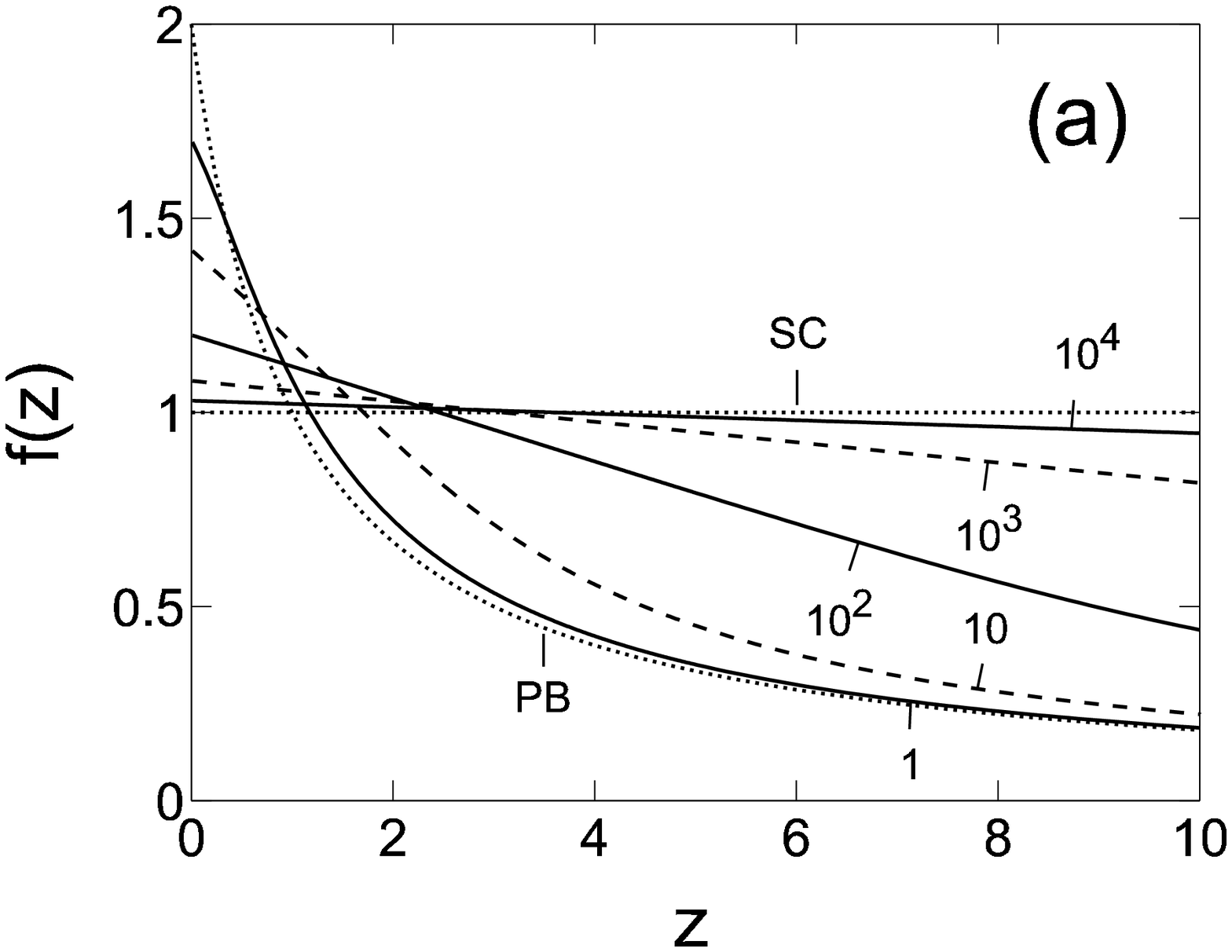}}
\hspace{1cm}
\scalebox{0.45}{\includegraphics{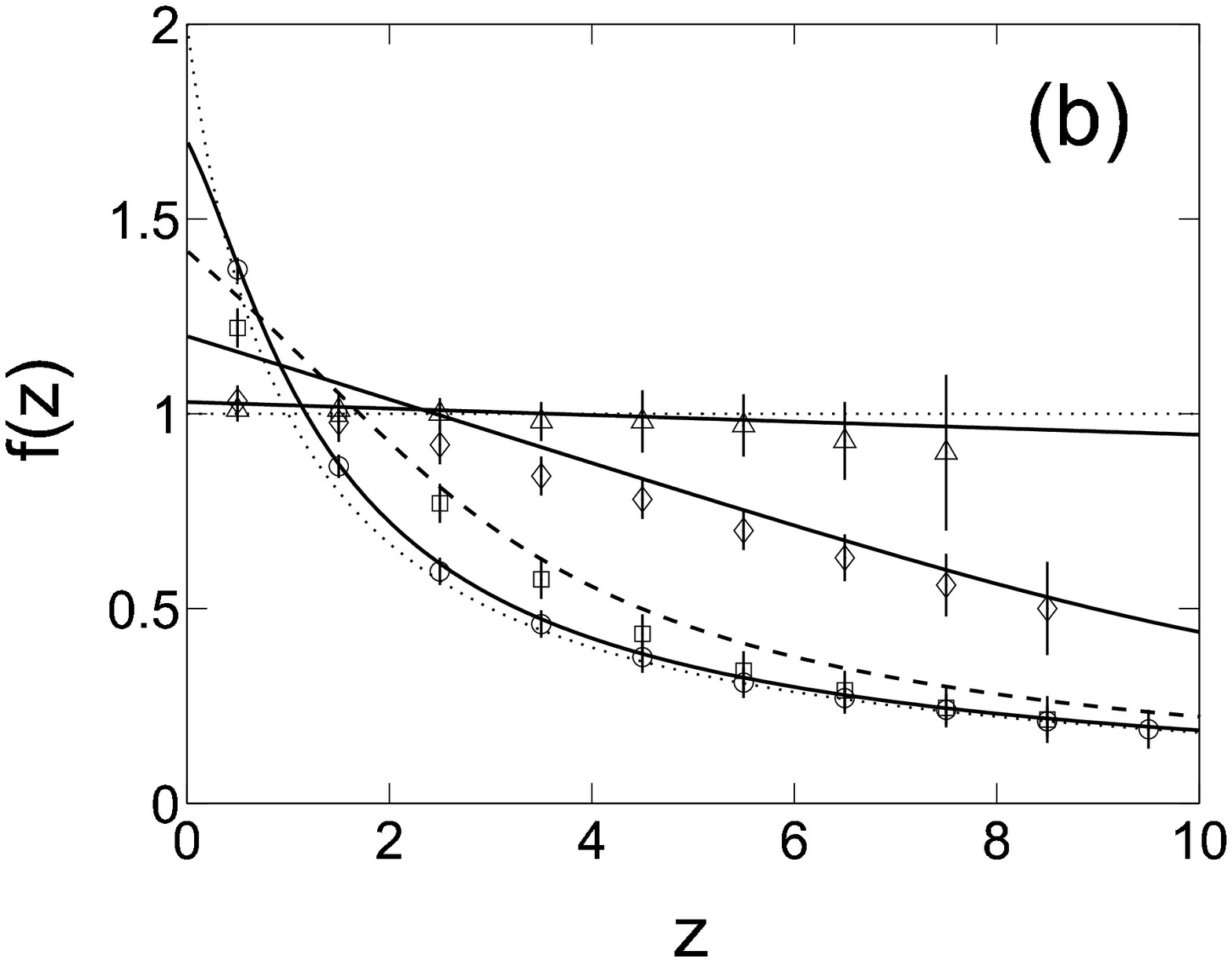}}
\caption{
(a) Electrostatic force acting on a test
charge, $f(z)$, numerically calculated using the mean field
equation (\ref{PBeq}).
Alternating solid and dashed lines show $f(z)$ for 
$\Xi = 1$, 10, $10^2$,$10^3$, and $10^4$. The dotted
lines show the PB and SC forms of $f(z)$, $f_{\rm PB}(z) = 1/(z+1)$ and 
$f_{\rm SC}(z) = {\rm exp}(-z)$. (b) Comparison of $f(z)$, calculated
from Eq.~(\ref{PBeq}) (solid and dashed lines, same as in 
part (a)), with results from Monte-Carlo simulation 
\cite{MoreiraPrivate}, adapted from 
Ref.~\cite{MoreiraNetz02} ($\Xi = 1$, circles; $\Xi = 10$, squares;
$\Xi=10^2$, diamonds; $\Xi = 10^4$, triangles). Values of $f(z)$ are
obtained from simulation results for $n(z)$ using the exact relation
${\rm d}{\rm log}n(z)/{\rm d}z = -f(z)$. Numerical estimation of
the derivative of ${\rm log}\left[n(z)\right]$ results in 
relatively large error bars, which are shown as vertical lines.
}
\end{figure*}
We consider first the behavior of $f(z)$,
defined in Eq.~(\ref{fofz}), close to the charged
plate. Fig.~3(a) shows this behavior for 
$\Xi$ = 1, 10, $10^2$, $10^3$ and $10^4$ (alternating
solid and dashed lines). The curves were obtained
from a numerical solution of the partial differential
equation (PDE), Eq.~(\ref{PBeq}) (see Appendix~\ref{ap:numerical}
for details of the numerical scheme). 
For comparison the weak coupling (PB) and strong coupling (SC)
limits are shown using dotted lines:
\begin{equation}
f_{\rm PB}(z) = \frac{2}{z+1} \ \ \ \ ; \ \ \ \ 
f_{\rm SC}(z) =  1
\end{equation}
As $\Xi$ increases $f(z)$ gradually shifts from PB to SC
behavior. At $\Xi = 10^{4}$, $f(z)$ is very close to $1$ within the range of
$z$ shown in the plot, although there is still a noticeable small deviation
from unity. 

In Fig.~3(b) these results are compared with simulation data
(symbols),
adapted from Ref.~\cite{MoreiraNetz02}. The value of $f(z)$ was 
obtained from the simulation results for $n(z)$ using the relation
${\rm d}{\rm log}[n(z)]/{\rm d}z = -f(z)$ 
\footnote{
The numerical differentiation
of ${\rm log}[n(z)]$ leads to large error bars because
$n(z)$, as obtained from the simulation, is noisy. In principle
$f(z)$ could be evaluated more accurately during the simulation 
run by direct use of 
Eq.~(\ref{fofz}), {\it i.e.} by averaging the electrostatic force
acting on ions as function of their distance from the plane. 
\label{footnote1}}. 
Qualitatively our results agree very well with simulation. 
Note especially the gradual decrease of $f(z)$ with increasing
$z$ for $\Xi=100$ (diamonds): this value of $\Xi$ is far
away from both the weak coupling and the strong coupling limits. 
The regime where $f(z)$ decreases linearly with $z$ will be further
discussed in Sec.~\ref{smallerz}.

It was previously conjectured \cite{Netz01}
that for all values of 
$\Xi$ the SC limit is valid close enough to the charged plane. 
We note, however, that at contact with the plane $f(z)$ is different
from unity at small and intermediate values of $\Xi$. 
Hence it is not very meaningful to define a region close
to the plane where the SC limit is valid,
unless $\Xi$ is very large. Values of $f(z)$, extracted from simulation
data in Fig.~3(b), suggest the same conclusion, {\it i.e.},
$f(z)$ does not approach unity at contact with the plane.
A more accurate measurement of $f(z)$ in the simulation
is desirable because the error bars, as obtained in Fig.~3(b),
are quite large.

\subsubsection*{Results for $n(z)$}

\begin{figure}
\scalebox{0.45}{\includegraphics{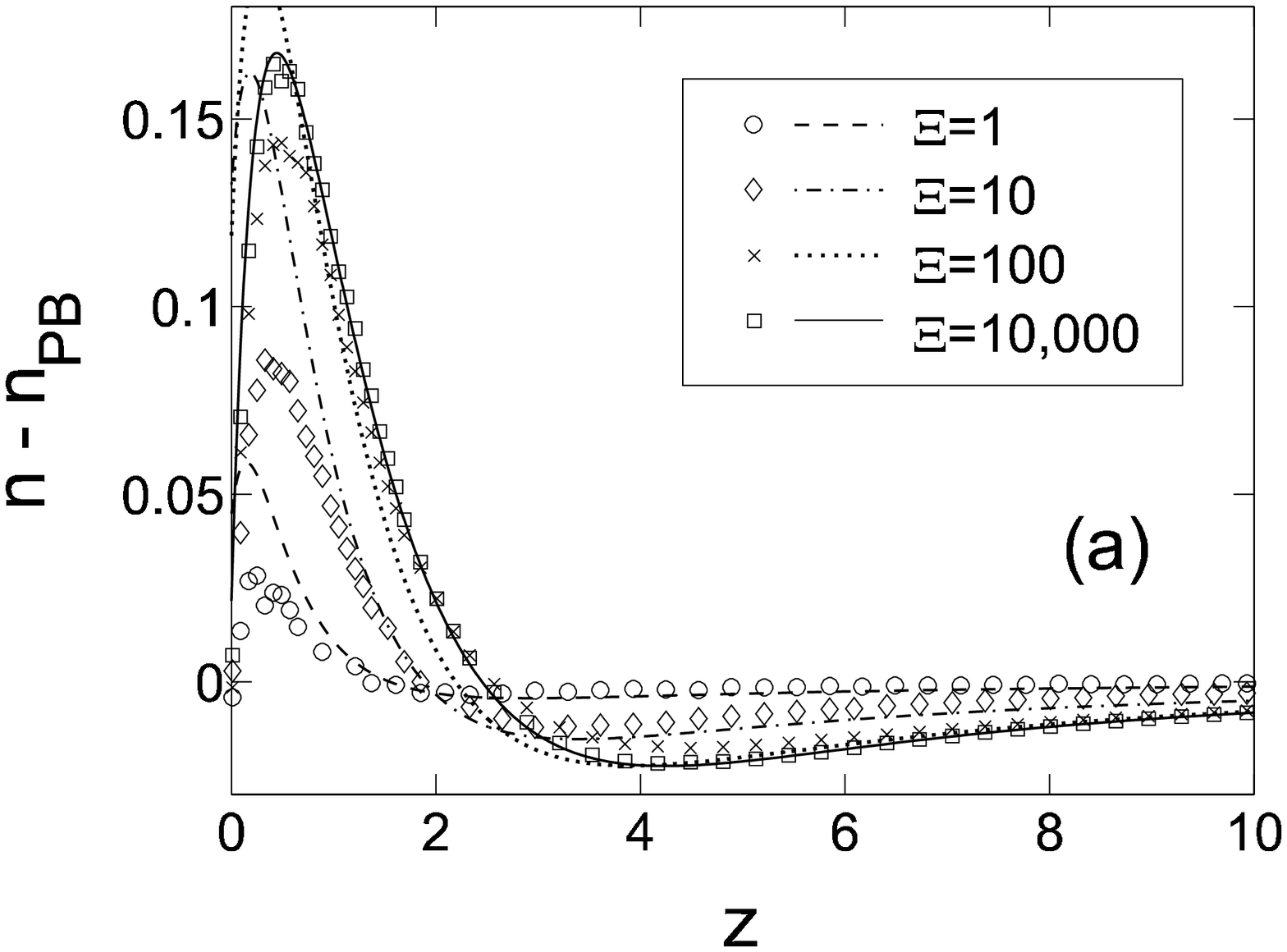}}
\hspace{1cm}
\scalebox{0.45}{\includegraphics{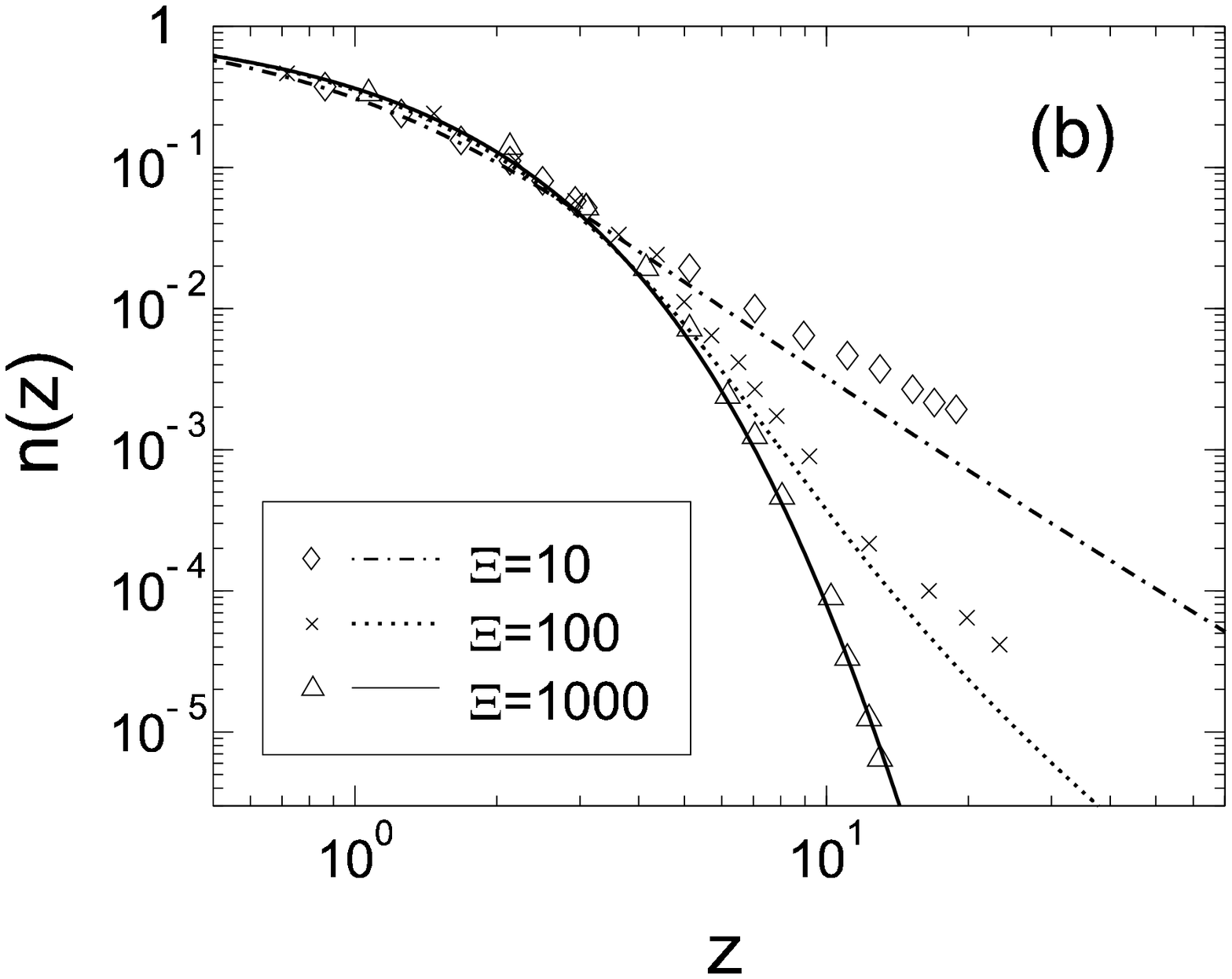}}
\caption{
(a) Correction to the PB density profile,
$n(z)-n_{\rm PB}(z)$, calculated numerically using the TCMF model,
as function of $z$ (lines). For comparison, symbols show the correction
obtained from Monte-Carlo simulation \cite{MoreiraNetz02,MoreiraPrivate}.
Four values of $\Xi$ are shown (see legend), 1, 10, $10^2$, and $10^4$.
(b) The density profile itself, $n(z)$, 
on a wider range of $z$ than in part (a) and
using logarithmic scales in both axes 
(lines - TCMF; symbols - MC simulation). 
}
\end{figure}
The density profile $n(z)$ can
be found numerically by integrating Eq.~(\ref{dlnrho1}) 
and use of the normalization condition (\ref{normalization})
\footnote{ 
It is also possible to find $n(z)$ using 
Eqs.~(\ref{model})-(\ref{modelnorm}), but integrating 
Eq.~(\ref{dlnrho1}) is numerically more accurate.}.
Figure~2 already demonstrated that $n(z)$ coincides with
the exact profiles, $n_{\rm PB}(z)$ and $n_{\rm SC}(z)$, 
in the limits of small and large $\Xi$.
Figure~4(a) shows the difference between $n(z)$ and
$n_{\rm PB}(z)$ for $\Xi = 1$, 10, 10$^2$, and
10$^4$, as calculated numerically in the TCMF model
(continuous lines). These results are compared with simulation
data (symbols)\cite{MoreiraNetz02,MoreiraPrivate}.

We first observe that the contact theorem is not obeyed in our
approximation:
$n(0)-n_{\rm PB}(0) = n(0)-1$ is different from zero.
This is an undesirable property, because
the contact theorem is an exact relation. 
The contact theorem is obeyed in the TCMF model
only in the limits of 
small and large $\Xi$, where the density profile as a whole 
agrees with the exact form, and the normalization condition
(\ref{normalization}) enforces $n(0)$ to be correct.
The violation at intermediate values of $\Xi$
is finite, small compared to unity, and peaks at $\Xi$ between 10 and
100. At these values of $\Xi$ the overall correction to PB is 
quite inaccurate 
at the immediate vicinity of the charged plate.
On the other hand, at $z$ larger than 1
our approximated results agree quite well with simulation
for all the values of $\Xi$, as
seen in Fig.~4(a). 

The violation of the contact theorem in the TCMF model
can be traced to a non-zero net force 
exerted by the ionic solution on itself (see Appendix~\ref{ap:contact}). 
This inconsistency results from the use of a mean field approximation
for the distribution of ions around the test charge, while the probability
distribution of the test charge itself is given by 
Eq.~(\ref{model}).

It is possible to evaluate exactly the first order correction
in $\Xi$ to the PB density profile in the TCMF model, the details
of which are given in Appendix~\ref{ap:smallxi}. 
This correction turns
out to be different from the exact first order 
correction, which was
calculated in Ref.~\cite{NetzOrland00} using a loop expansion
up to one loop order (also shown in the Appendix).
It is important to note, 
however, that the latter correction
provides a useful result only for relatively small values of $\Xi$.
At $\Xi$ of order 10 and larger TCMF results are much closer to 
simulation than those of the loop expansion. 

Further comparison with simulation is shown in Fig.~4(b). Here we
show the density $n(z)$ itself, rather than the difference with respect
to $n_{\rm PB}(z)$. The data is shown on a larger range
of $z$ than in part (a) and a logarithmic scale is used 
in order to allow small values of $n(z)$ to be observed far away from
the plate. In the range shown the TCMF model agrees semi-quantitatively
with simulation.

As a summary of this section we can say that the test charge mean
field (TCMF) model captures the essential behavior of 
the ion distribution at close and
moderate distances from the charged plate. Furthermore, all the 
available data from simulation agrees qualitatively with our
approximation's predictions. 

%===================================================================
\section{TCMF results far away from the charged plate}
%===================================================================
\label{sec:far}

\begin{figure*}
\scalebox{0.45}{\includegraphics{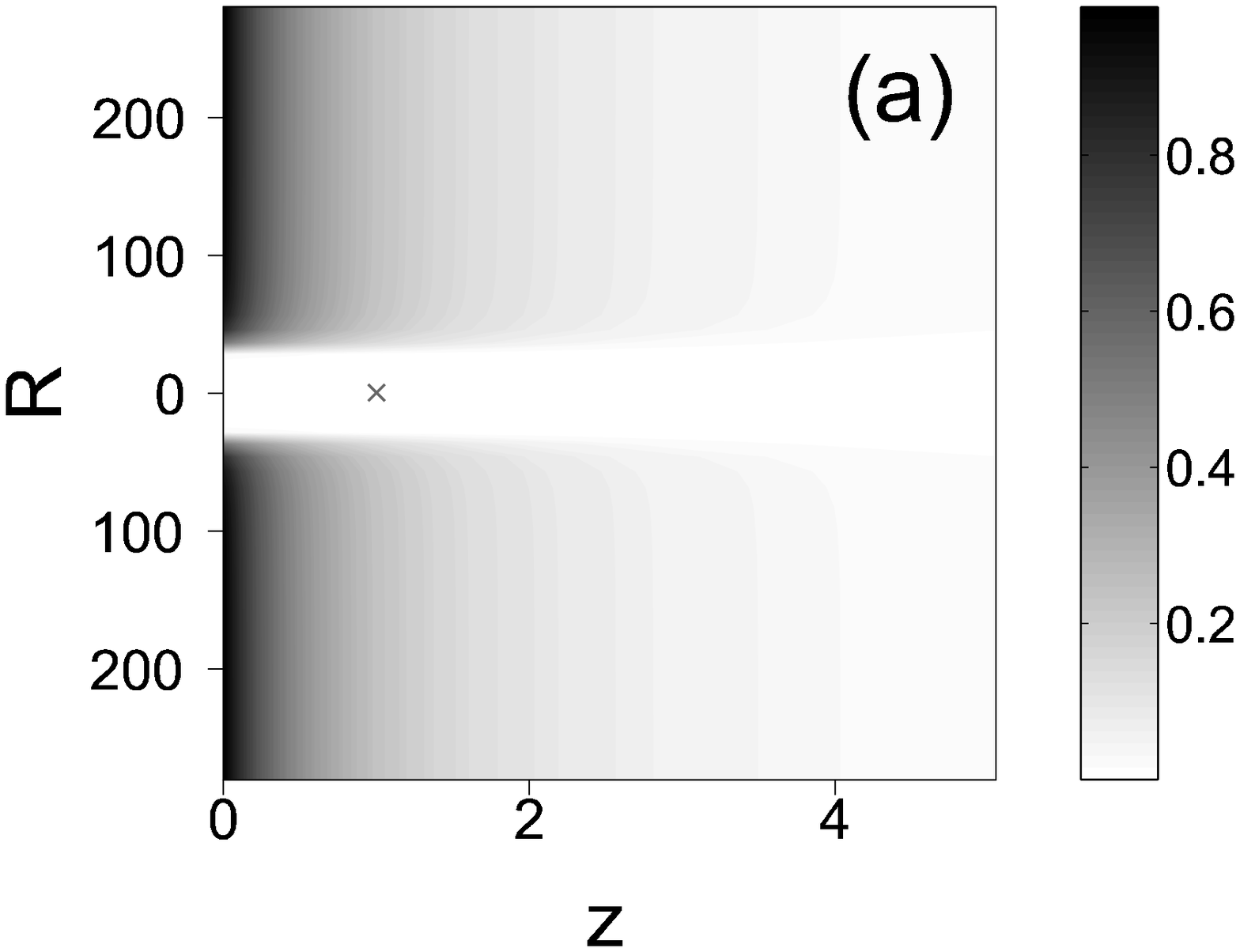}}
\hspace{1cm}
\scalebox{0.45}{\includegraphics{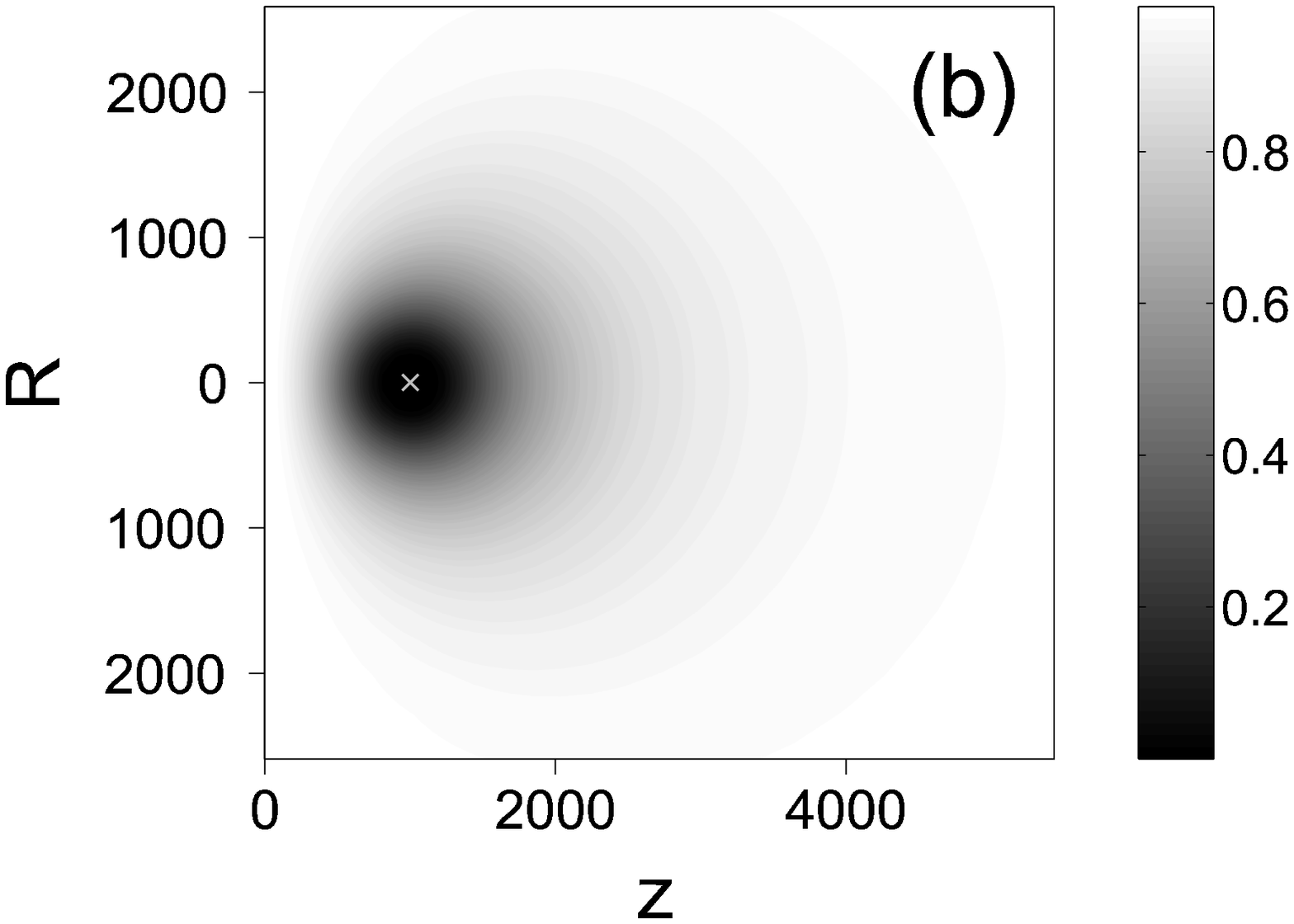}}
\caption{
(a) Scaled density of ions around a test charge,
positioned at $z_0 = 1$,
as obtained from Eq.~(\ref{PBeq}). The cross designates the
position of the test charge. In cylindrical coordinates the density
is a function only of $z$ (horizontal axis) and $r$ (vertical axis). 
Darker shading in the plot
means larger density (see also the legend on the right). The coupling
constant is $\Xi$ = 1,000. For $r$ larger 
than $\sqrt{\Xi}$ the profile, as function of $z$, 
quickly converges to the PB profile,
$n_{\rm PB}(z)$ (b) A similar plot as in part (a), but the test charge is 
now at $z_0 = \Xi$ = 1,000. Here the ratio between the density and the 
PB profile is shown, rather
than the density itself. This ratio is everywhere a number between 
zero (black) and one (white).
The effect of the test charge on the ion distribution is large only
within a correlation hole around the test charge,
having approximately a spherical shape and a
radius of order $\Xi$.
}
\end{figure*}

Our analysis of the ion distribution far away from
the charged plate is done, at first, strictly
within the context of the TCMF model, 
while a discussion of its relevance
to the exact theory is deferred to Sec.~\ref{sec:further}.
The main question of interest is whether a transition to PB behavior
occurs at sufficiently large $z$, even for large values of $\Xi$.

As a first step we will identify the 
important length scales characterizing the density distribution. 
Let us concentrate first on the
size of the correlation hole around a test charge.
Naively we may expect this size to be of order $\Xi$, 
due to the form of the bare potential, $\Xi/|{\bf r}-z_0\hat{\bf z}|$. 
A simple argument shows, however, that when the test charge is 
close to the charged plane the size of the correlation hole is much 
smaller than $\Xi$. Assume, roughly, that the mobile ion density
is zero within a cylindrical shell of radius $R$ around the test charge. 
The amount of charge depleted from this cylinder is then
equal to $R^2/2$, since the surface charge per unit area is
equal to $1/{2\pi}$ (see Eq.~(\ref{PBeq})). 
This depleted charge must balance exactly the charge of the test 
particle, equal to $\Xi$, yielding a 
cylinder radius that scales as $\sqrt{\Xi}$ rather than as $\Xi$. 

In order to put this argument to test,
Fig.~5(a) shows the density of mobile ions calculated from Eq.~(\ref{PBeq})
with a test charge at $z_0 = 1$, having $\Xi$ = 1,000. 
The shape of the correlation hole is roughly cylindrical and its radius is
indeed of order $\sqrt{\Xi} \simeq 30$. 
The influence of the test charge on its 
surroundings is very non-linear, with a sharp spatial transition from the
region close to the test charge, where the density is nearly zero,
to the region further away, 
where the effect of the test charge is very small.
At larger separations from the
plate the qualitative picture remains the same, as long as $z_0$ is small 
compared to $\sqrt{\Xi}$ and provided that $\Xi \gg 1$. 

A very different distribution of mobile ions 
is found when $z_0$ is of order $\Xi$, as shown in 
Fig.~5(b). The coupling parameter is the same as in part (a), 
$\Xi$ = 1,000, but the test charge is now at $z_0$ = 1,000. 
Instead of showing directly the density of
mobile ions as in part (a), the figure shows the 
ratio between this density and $n_{\rm PB}(z) = 1/(z+1)^2$.
This ratio 
is now very close to unity near the charged 
plane, where most of the ions are present. It is small compared to 
unity only within a spherical correlation hole around the test charge,
whose size is of order $\Xi$. 

The above examples lead us to divide our discussion of
the $z$ dependence into two regimes:

\subsection{$z < \sqrt{\Xi}$}
\label{smallerz}

In order to justify use of the cylindrical correlation hole approximation 
within this range, let us assume such a correlation hole and calculate
the force acting on the test charge:
\begin{equation}
f(z) \simeq \int_0^{\infty} {\rm d}z'\, n_{\rm PB}(z') 
\frac{z'-z}{\sqrt{\Xi + (z'-z)^2}}
\label{fapp}
\end{equation}
where $n_{\rm PB}(z')$ is given by Eq.~(\ref{npb}), the radius of the
cylindrical region from which ions are depleted is taken as $\sqrt{\Xi}$,
and the expression multiplying $n_{\rm PB}(z')$
is the force exerted by a charged sheet having a 
circular hole of radius $\sqrt{\Xi}$ and positioned in the
plane $z'$.
Figures 6 (a) and (b) show a comparison of this approximation
(dashed lines) with that obtained from a full numerical
solution of Eq.~(\ref{PBeq}) 
(solid lines). The coupling parameter 
 is equal to 10,000 in (a) and to 100 in (b). In
both cases the approximation works well up to $z_0 \simeq \sqrt{\Xi}$.
In Fig.~6 (c) the force acting on a test charge at contact
with the plane, $f(0)$, is shown for five values of $\Xi$ between 1 and 
10,000 (symbols), and compared with the approximation of
Eq.~(\ref{fapp}) (solid line). Note that
Eq.~(\ref{fapp}) is not a good approximation when
$\Xi$ is of order unity or smaller, 
since the correlation hole is then small 
compared to the width of the ion layer.

\begin{figure}
\scalebox{0.35}{\includegraphics{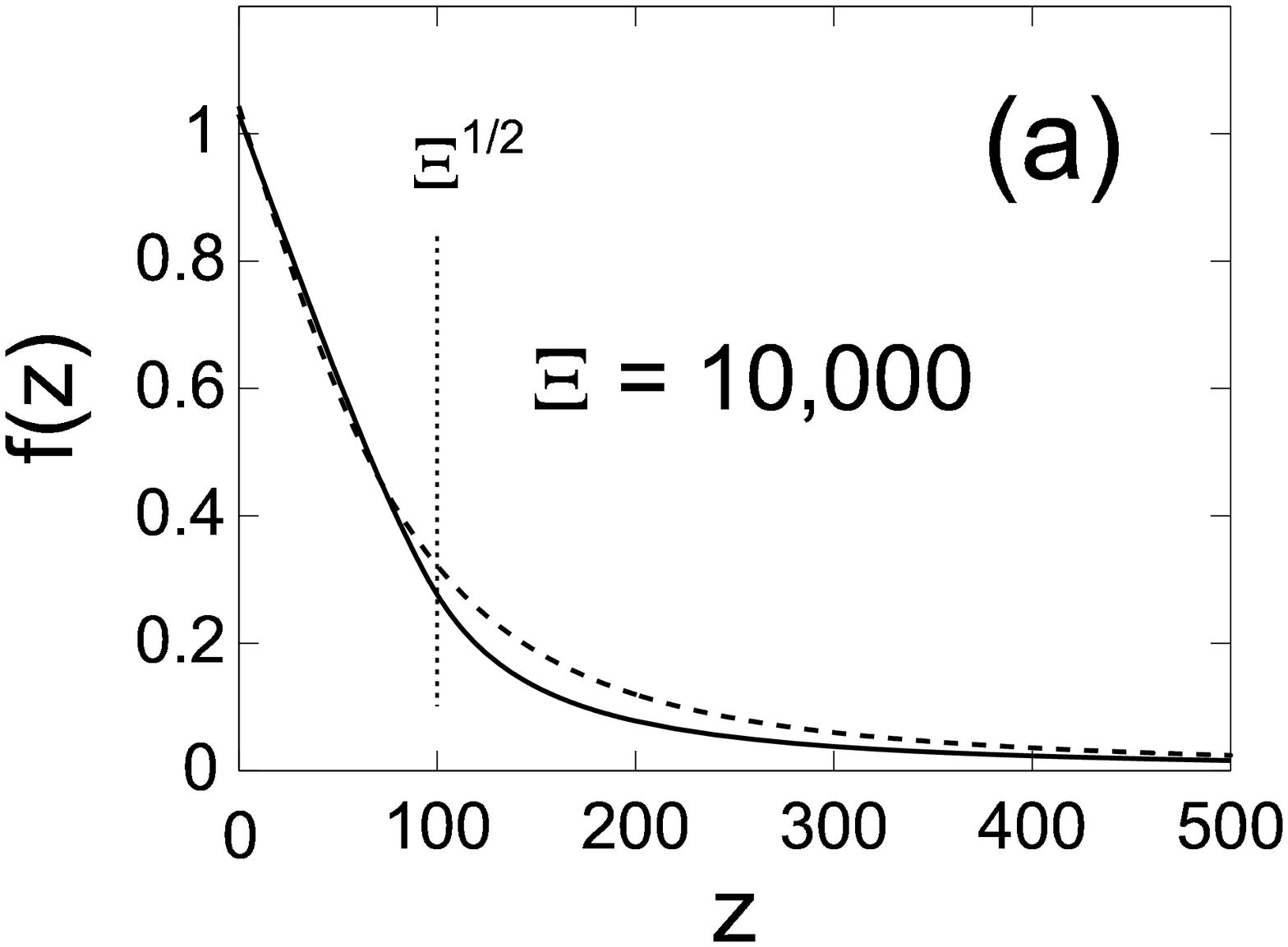}} \\
\vspace{0.3cm}
\scalebox{0.35}{\includegraphics{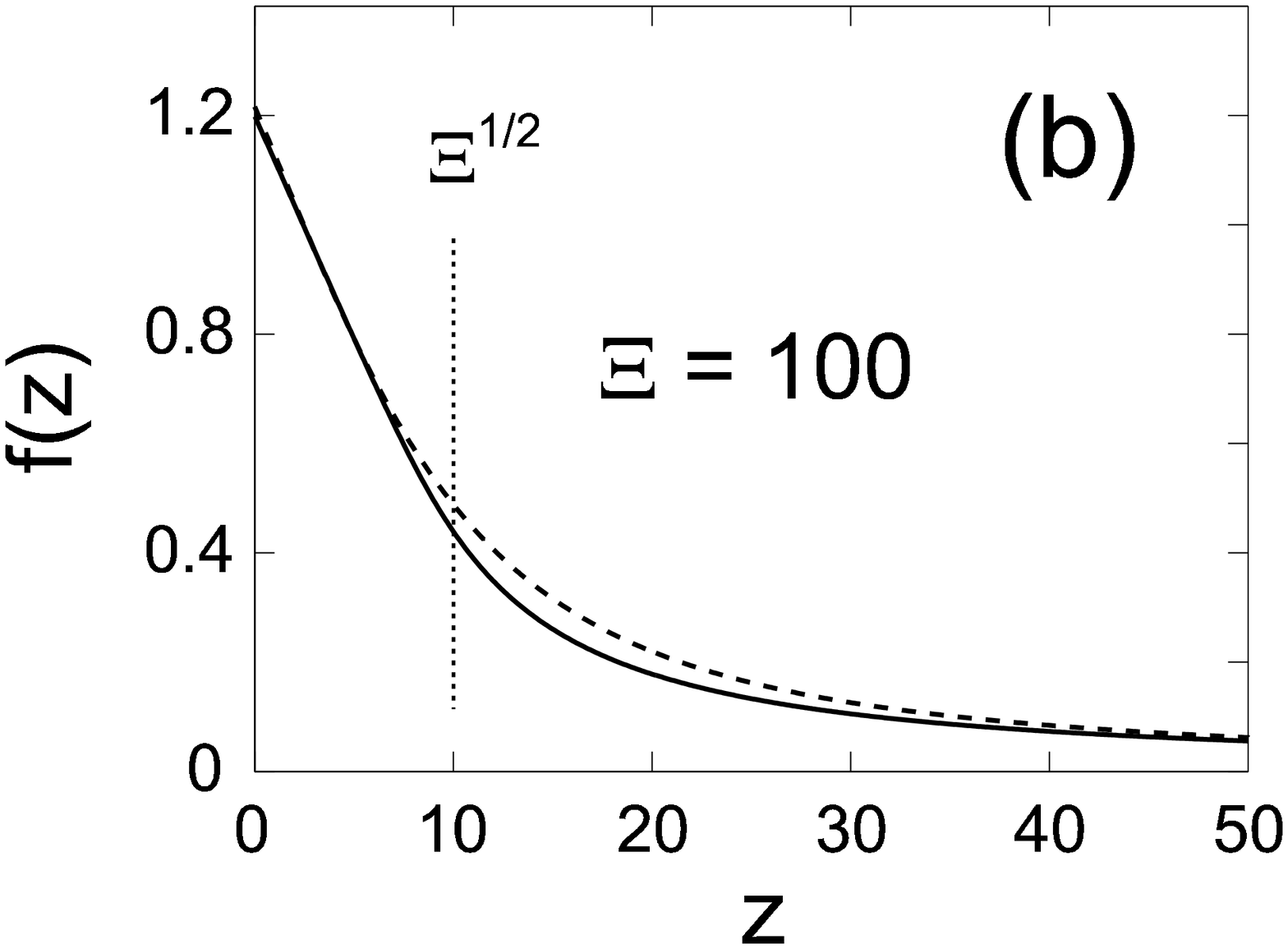}} \\
\vspace{0.3cm}
\scalebox{0.35}{\includegraphics{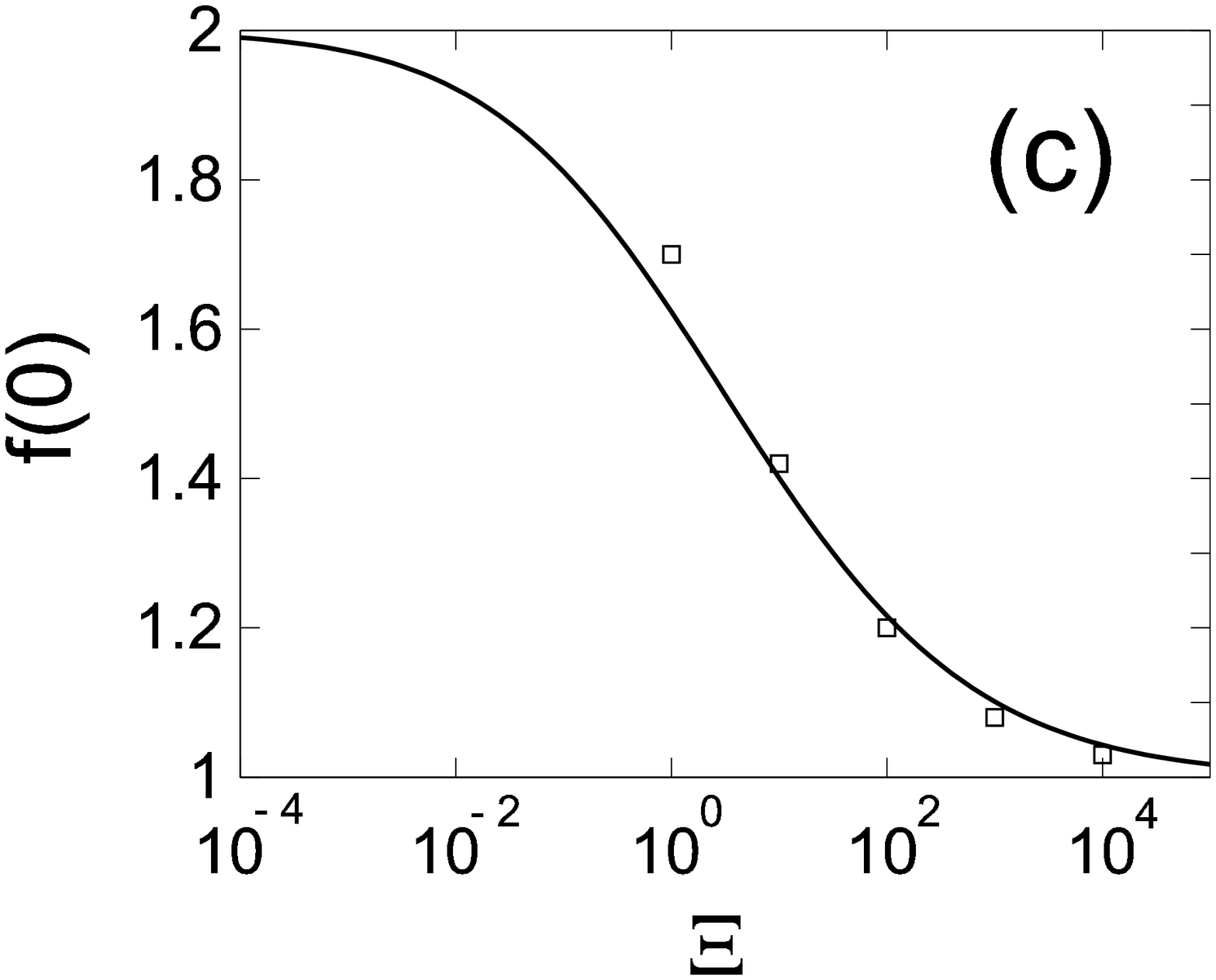}}
\caption{
(a), (b) 
Comparison of the approximation to $f(z)$
given by Eq.~(\ref{fapp}) (dashed line), 
with a full numerical solution of the PDE (solid line). The coupling
parameter $\Xi$ is equal to 10,000 in (a) and to 100 in (b). Note that
the approximation shown in the dashed line is good up to a distance
from the plate equal to about $\sqrt{\Xi}$ in both cases. A distance
of $\sqrt{\Xi}$ from the charged plate is designated by the vertical 
dotted lines. Part (c)
shows a comparison of 
$f(0)$ in the approximation given by Eq.~(\ref{fapp})
(symbols) and in the exact PDE
solution (solid line) for a wide range of $\Xi$ values.
}
\end{figure}

\subsection{$z > \sqrt{\Xi}$}

When the test charge is far away from the plane, its effect
close to the charged plate is small, suggesting that a
linear response calculation may be applicable:
\begin{equation}
f(z) = f_{\rm PB}(z) + \Xi f_1(z)
\label{flin}
\end{equation}
The first term in this equation is the PB value of $f(z)$, 
while $f_1(z)$ can be calculated
using previous results of Ref.~\cite{NetzOrland00}:
\begin{eqnarray}
f_1(z) & = & \frac{1}{2}
\frac{{\rm d}g(z)}{{\rm d}z} = \frac{1}{4(z+1)^3} 
\times {\Big \{} 8z \nonumber \\
& - & (1+i){\rm e}^{(1-i)z}
\left[1-z+(1-2i)z^2+z^3\right] \nonumber \\
& & \times E_1\left[(1-i)z\right] \nonumber \\
& - & (1-i){\rm e}^{(1+i)z}\left[1-z+(1+2i)z^2+z^3\right]
\nonumber \\
& & \times E_1\left[(1+i)z\right]{\Big \}}
\label{f1}
\end{eqnarray}
where $g(z)$ was defined in Ref.~\cite{NetzOrland00} 
and is given by Eq.~(\ref{gofz}), and 
$E_1(z)$ is the exponential-integral function
\cite{AStegun}. 
We prove the first equality of Eq.~(\ref{f1}) in 
Appendix~\ref{ap:smallxi}. 
Figure 7 shows $f_1(z)$ (solid line) together with
its asymptotic form for large $z$ (dashed line),
\begin{equation}
f_1(z) \simeq \frac{3}{4z^2} \ \ \ , \ \ \ z \gg 1
\label{flinapp}
\end{equation}
Note that the asymptotic form of $\Xi f_1(z)$ has the same $z$ 
dependence as the
electrostatic force exerted
by a metallic surface, equal to $\Xi/(4z^2)$ in our notation, 
but the numerical prefactor is different.
\begin{figure}
\scalebox{0.45}{\includegraphics{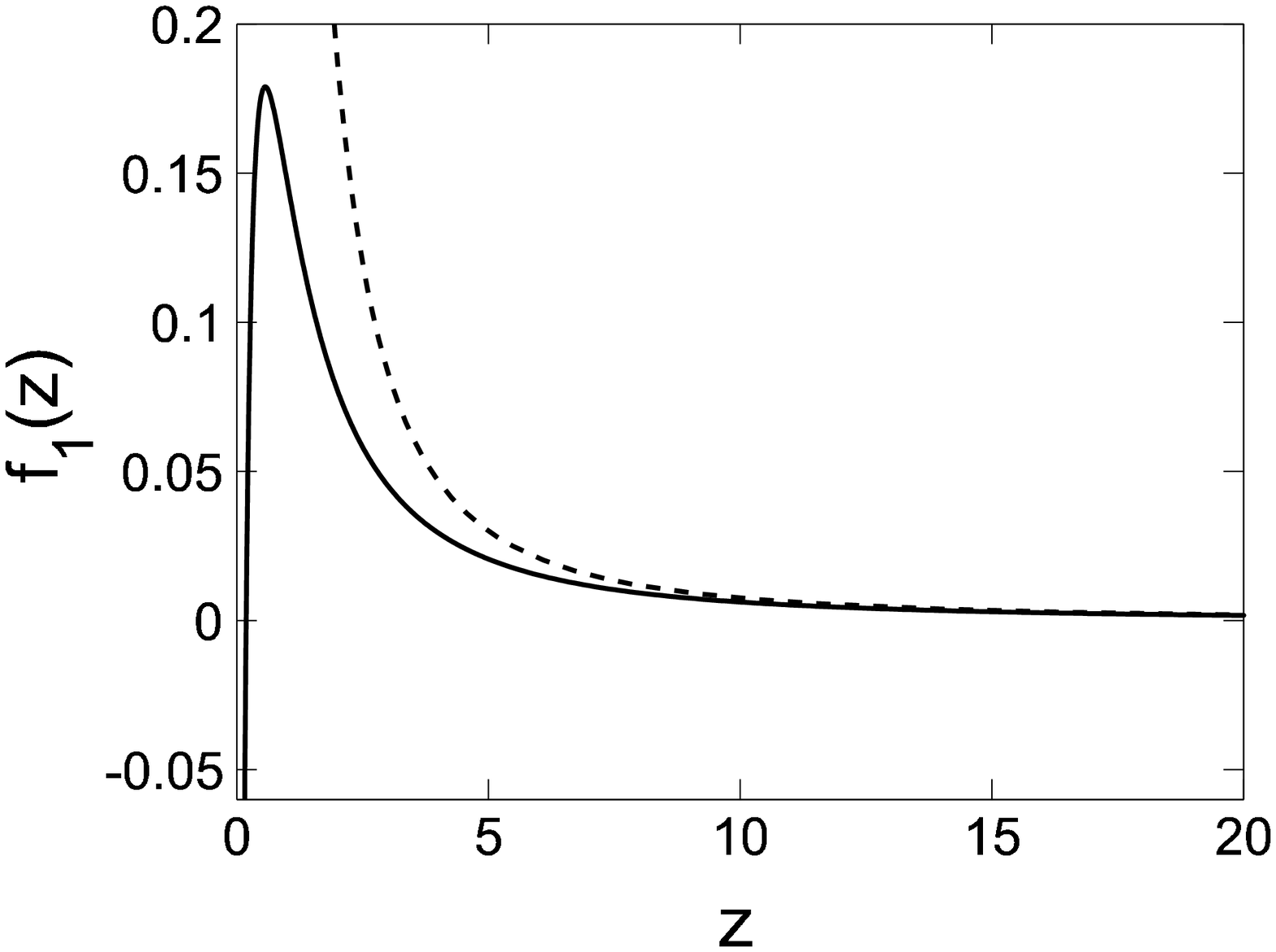}}
\caption{
First order (linear) term in an expansion
of $f(z)$: $f(z) = f_{\rm PB}(z) + \Xi f_1(z) + \ldots$, 
Eq.~(\ref{f1}), as obtained from
the loop expansion of Ref.~\cite{NetzOrland00}.
The dashed line shows the asymptotic form of $f_1(z)$
at large $z$, $f_1(z) \simeq 3/(4z^2)$.
}
\end{figure}

Although the influence of the test charge is small near the surface, its
influence on ions in its close vicinity is highly 
non-linear and definitely not small. 
Hence the applicability of Eq.~(\ref{flin}) 
is far from being obvious when $\Xi$ is large. We check this numerically by
calculating $f(z)-f_{\rm PB}(z)$ and comparing
with $\Xi f_1(z)$. The results are shown in Fig.~8(a), for 5 values
of $\Xi$: 1, 10, $10^2$, $10^3$, and $10^4$.

\begin{figure*}
\scalebox{0.45}{\includegraphics{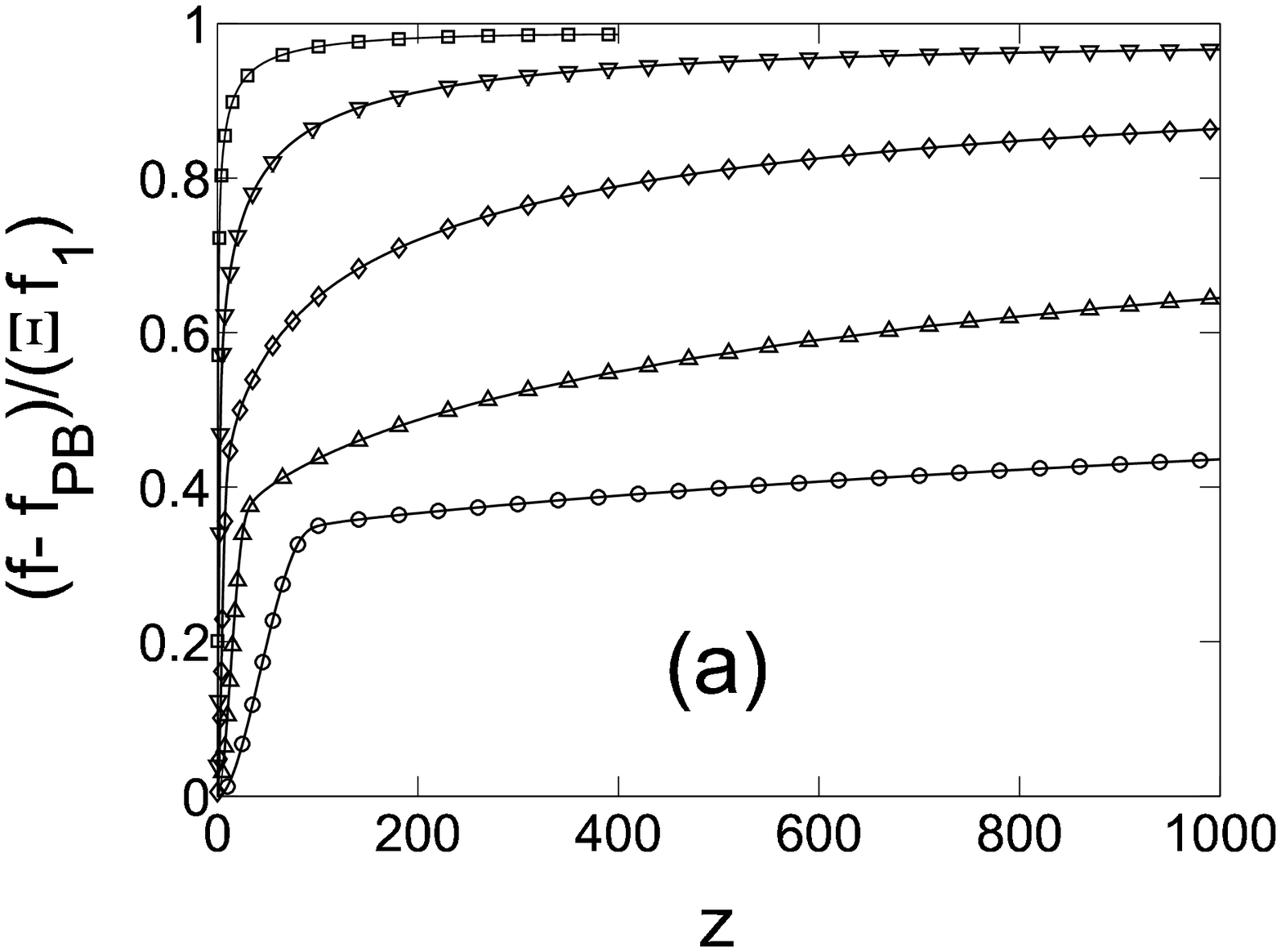}}
\hspace{0.5cm}
\scalebox{0.45}{\includegraphics{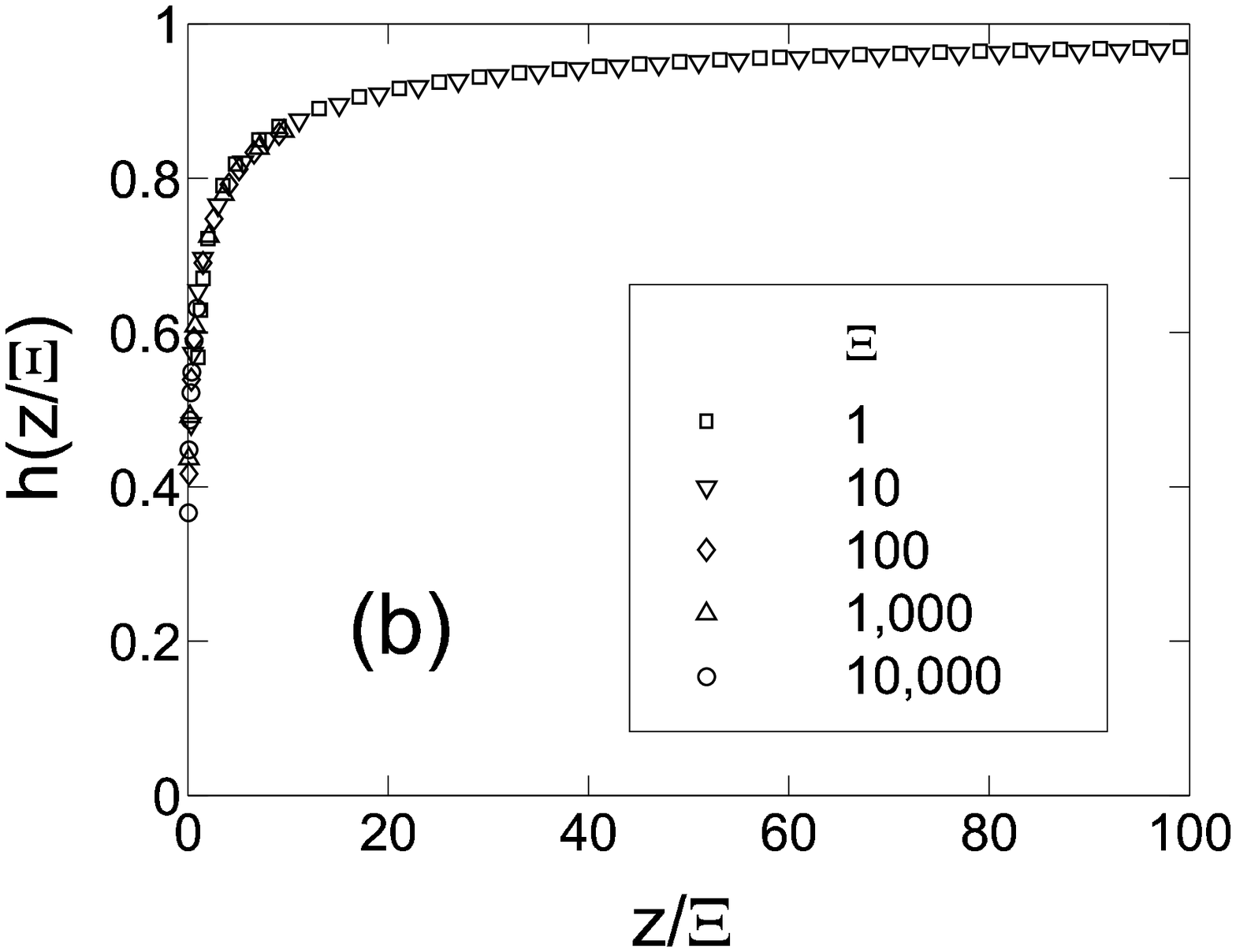}}
\caption{
Comparison between the correction to $f(z)$
relative to $f_{\rm PB}(z)$, with the linearized
expression $\Xi f_1(z)$. In (a) the ratio $[f(z)-f_{\rm PB}(z)]/[\Xi f_1(z)]$
is shown as function of $z$ for five different values of $\Xi$:
1, 10, 10$^2$, 10$^3$, and 10$^4$ (see legend in part (b); symbols show
the same quantity as the solid line and are displayed in order to 
distinguish between the five lines). The ratio approaches unity at 
$z$ much larger than $\Xi$.
In (b) the same data as in (a) is shown as function of $z/\Xi$, leading
to an almost perfect collapse of the five data sets on a single curve.
}
\end{figure*}

For each value of $\Xi$ the ratio $(f-f_{\rm PB})/(\Xi f_1)$ (shown in
the plot) approaches unity as $z$ is increased, showing that
Eq.~(\ref{flin}) does become valid at sufficiently large $z$. The 
approach is, however, rather slow: a value close to unity is reached only when
$z \gg \Xi$. At $z = \Xi$ the ratio is approximately equal to 0.6 
in all five cases. We conclude that the linear 
approximation of Eq.~(\ref{flin}) is applicable only for $z \gg \Xi$. 
Note that at these
distances from the charged plate the linear correction itself is
very small compared to the PB term,
\begin{equation}
\frac{\Xi f_1(z)}{f_{\rm PB}(z)} \simeq \frac{3\Xi}{4z^2}\frac{z+1}{2} 
\simeq \frac{3}{8}\frac{\Xi}{z} \ll 1
\end{equation} 
where we also assumed that $z \gg 1$ and used Eq.~(\ref{flinapp}).

Further insight on the results shown in Fig.~8(a) is obtained by noting
that all of them approximately
collapse on a single curve after scaling the $z$ coordinate by $\Xi$. 
This curve, denoted by $h(z/\Xi)$,
is shown in Fig.~8(b): 
\begin{equation}
f(z)-f_{\rm PB}(z) \simeq \Xi f_1(z) \times h\left(\frac{z}{\Xi}\right)
\label{collapse}
\end{equation}
In order to demonstrate at what range of $z$ values
this scaling result is valid the same data
is shown in Fig.~9 using a logarithmic scale in the horizontal ($z/\Xi$)
axis. It is then seen clearly that 
(\ref{collapse}) holds for $z/\Xi$ larger than a minimal value, 
which is proportional to $\Xi^{-1/2}$. 
The vertical arrows designate $z/\Xi = 1.5/\sqrt{\Xi}$ for each
value of $\Xi$, 
approximately the point where the scaling becomes valid.
Returning to consider $z$ itself, we conclude that (\ref{collapse}) holds
for $z \gtrsim 1.5 \sqrt{\Xi}$. This result justifies the separation of
the $z$ dependence into two regimes, $z < \sqrt{\Xi}$ and
$z > \sqrt{\Xi}$.

\begin{figure}
\scalebox{0.45}{\includegraphics{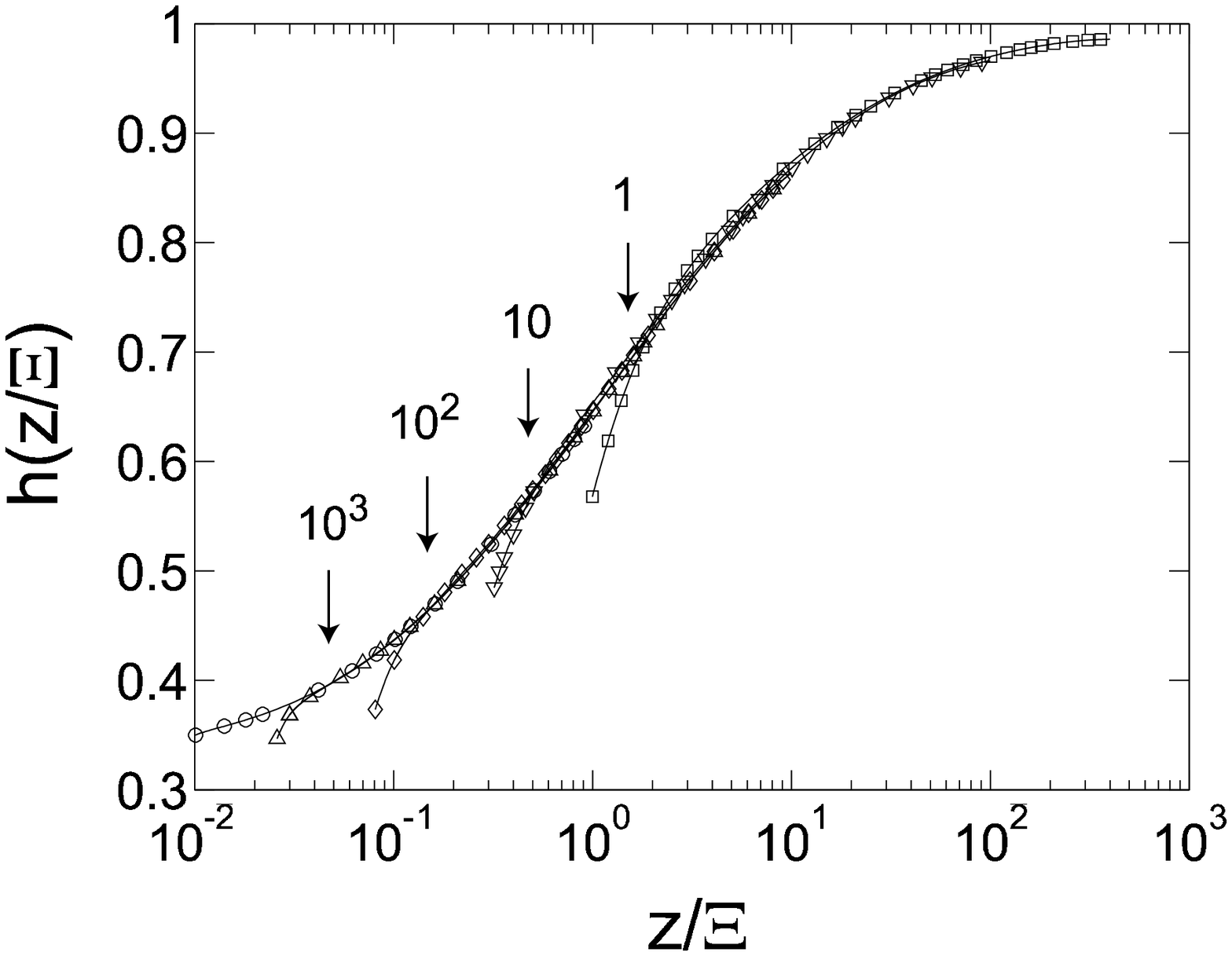}}
\caption{
Same data as in Fig.~9(b), shown using
a logarithmic scale in the horizontal (z/$\Xi$) axis. The approximated
collapse of the different data sets, corresponding to different values of 
$\Xi$, is seen to be valid only in the regime 
$z \gtrsim \sqrt{\Xi}$. The vertical
arrows mark $z = 1.5 \sqrt{\Xi}$ for $\Xi = 1$, 10, 10$^2$, and 10$^3$,
above which 
the scaling (\ref{scaling}) is approximately valid.
}
\end{figure}
\begin{figure}
\scalebox{0.45}{\includegraphics{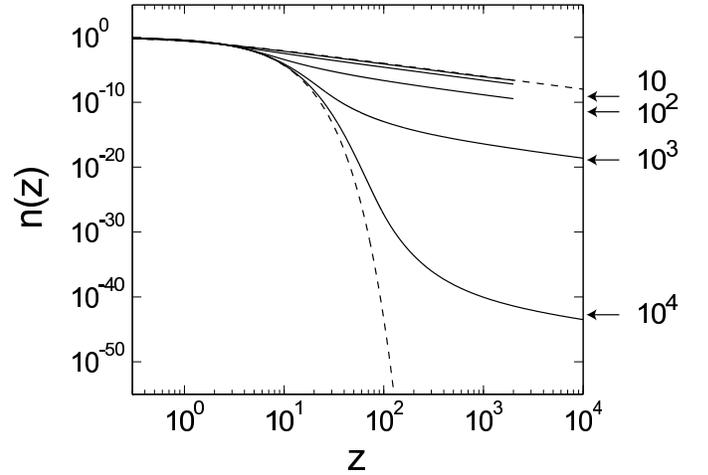}}
\caption{
Scaled ion density, $n(z)$ calculated
using the TCMF model, shown for five
different values of $\Xi$ (solid lines, top to bottom): 
1, 10, 10$^2$, 10$^3$, and 10$^4$. A logarithmic
scale is used on both axis, allowing the behavior far away from the charged
plate to be seen. The dashed lines show $n_{\rm PB}(z)$ (upper dashed line) 
and $n_{\rm SC}(z)$ (lower dashed line). 
At $z \gg \Xi$ the density profile is proportional 
to $n_{\rm PB}(z)$, with a prefactor whose logarithm scales as $\sqrt{\Xi}$.
To demonstrate this the horizontal arrows mark the value of 
${\rm exp}(-0.8 \sqrt{\Xi}) \times
n_{\rm PB}({10^4})$ for $\Xi = 10$, $10^2$, $10^3$, and $10^4$.
The prefactor 0.8 is an approximate fitting parameter.
}
\end{figure}
We finally turn to consider the ion density $n(z)$. Using 
Eqs.~(\ref{flinapp}) and (\ref{collapse}) we can write
\begin{equation}
-\frac{\rm d}{{\rm d}z}n(z) = f(z) \simeq f_{\rm PB}(z) + \frac{1}{\Xi}
\cdot \frac{3}{4}\left(\frac{z}{\Xi}\right)^{-2} h\left(\frac{z}{\Xi}\right)
\end{equation}
which leads, upon integration, to the approximate scaling result,
\begin{equation}
n(z) \simeq c_0(\Xi) \frac{1}{(z+1)^2}\eta\left(\frac{z}{\Xi}\right)
\label{nscaling}
\end{equation}
where
\begin{equation}
{\rm log}[\eta(u)] = -\frac{3}{4}\int_u^\infty{\rm d}u'\,\frac{h(u')}{u'^2}
\end{equation}
The prefactor $c_0(\Xi)$ depends, through the normalization condition,
on the behavior of $n(z)$ close to the charged plane where the scaling
form of Eq.~(\ref{nscaling}) is not valid. Equation 
(\ref{nscaling}) is indeed validated by the 
numerical data and applies for $z \gtrsim \sqrt{\Xi}$ and $\Xi \gg 1$.

The density itself, $n(z)$, is plotted in Fig.~10
using a semi-logarithmic scale, for $\Xi$ = 1, 10, 10$^2$,
10$^3$, and $10^4$.
At $z \gg \Xi$ $n(z)$ is proportional to $n_{\rm PB}(z)$,
as expected from Eq.~(\ref{nscaling}).
The prefactor $c_0$ is extremely small for large $\Xi$. 
We recall that $c_0$ is mainly determined by the behavior 
close to the charged plane, where $f(z)$ is of order unity up to 
$z \lesssim \sqrt{\Xi}$. Following this observation we can 
expect ${\rm log}[c_0(\Xi)]$ to scale roughly as
$\sqrt{\Xi}$. This estimate is validated by the numerical
results and is demonstrated in the figure using the horizontal arrows.
For $\Xi = 10$, $10^2$, $10^3$, and $10^4$ these arrows show an estimate 
for $n(z)$ at $z = 10^4$, given by
\begin{displaymath}
n_{\rm PB}(z)\times {\rm exp}(-0.8\sqrt{\Xi})
\end{displaymath}
in very good agreement with the actual value of $n$.

%=============================================================
\section{Further Discussion}
%=============================================================
\label{sec:further}

At this point we may ask to what extent our results represent 
the behavior of $n(z)$ in the exact theory. Before discussing this
question we turn our attention for a moment to the system that
our problem was mapped into in Eq.~(\ref{PBeq}) -- 
that of a single ion of
valence $\Xi$ in contact with a charged plane and its monovalent counterions.
The results of Sections \ref{sec:numerical} and \ref{sec:far} 
can be regarded as exact for such a system, 
provided that the monovalent ions
are weakly correlated (having, by themselves, a small coupling 
parameter as determined from their charge and that of the planar surface). 
These results are thus of direct relevance to the interaction
of a large multivalent macroion with a charged surface that is
immersed in a weakly correlated solution of counterions.

Returning to the original question, we separate our discussion
according to the scaling results of the numerical analysis:

\subsection{$z < \sqrt{\Xi}$}

When $\Xi$ is very large 
a test charge at $z < \sqrt{\Xi}$ is essentially decoupled from the rest 
of the ionic solution, feeling only the force exerted 
by the charged plane. This is the reason why an
exponential decay, $n(z) \sim {\rm exp}(-z)$, is obtained
in our model, as well as in simulation and in the
perturbative strong coupling expansion of Ref.~\cite{Netz01}.
However, at intermediate values
of $\Xi$ such as 10, 100, or 1000 our results show that
this exponential decay is 
only a rough approximation.
The decoupling of a test charge from the rest of the ions is only 
partial, even at $z = 0$, leading
to a value of $f(z)$ that is (\textit{i}) larger than unity at $z = 0$ and 
(\textit{ii}) considerably
smaller than unity at $z = \sqrt{\Xi}$. Both of these predictions are
validated by simulation, as shown in Fig.~3(b). 

The quantitative agreement in $f(z)$ between our model and
simulation is 
surprisingly good, considering
that the ionic environment surrounding the test charge is different
in our approximation, compared to the exact theory. This good agreement
can be attributed to the correct length scales that characterize
the approximate ionic environment: $\sqrt{\Xi}$ in the
lateral direction, and 1 in the transverse direction. Indeed, in the
lateral direction our results can be compared with pair distributions
that were obtained in Monte-Carlo simulation \cite{MoreiraNetz02}.
The pair distributions found in the simulation
are characterized by a strong depletion within
a correlation hole having diameter of order $\sqrt{\Xi}$, in great 
similarity to Fig.~5(a). 
What is not captured by our approach is that multiple maxima
and minima exist at $\Xi \gtrsim 100$ 
\footnote{
In principle, a more accurate evaluation of the pair distribution
function, possibly capturing its oscillatory nature, may be
obtained from an approach similar to the TCMF with two 
test charges instead of one.
}. 
Nevertheless, even at the very large coupling parameter value $\Xi=10^4$
these oscillations decay quite rapidly with lateral distance, and we 
can still say that the
TCMF model captures the most significant feature of the pair
distribution (namely, the structure of the correlation hole).

\subsection{$z > \sqrt{\Xi}$}

Throughout most of this section we concentrate on the case
$z > \Xi$, while a short discussion of the range 
$\sqrt{\Xi} < z < \Xi$ is presented at the end of this section.

Our model predicts a transition to algebraic decay of
$n(z)$ at $z \gtrsim \Xi$. Similar predictions were made in 
Ref.~\cite{Shklovskii} and in Ref.~\cite{MoreiraNetz02}, where
it was estimated that mean field results are valid
for $z > \Xi {\rm log}\Xi$ based on
a perturbative expansion around mean field. There
are currently no available results from simulation in this regime.

A mean field behavior is obtained in our model
in the sense that
\begin{equation}
f(z) \simeq f_{\rm PB}(z) = \frac{1}{z+1}
\label{fasym}
\end{equation}
decays as $1/z$ for large $z$. Nevertheless
the finer details in our results do not match the form predicted
by PB theory. The starting point for the following discussion
is an hypothesis that sufficiently far from the plate the 
exact density follows a PB form,
\begin{equation}
n_{\rm asym}(z) = \frac{1}{(z+b)^2}
\label{nasym}
\end{equation}
where $b$ (or $\mu b$ in the original, non-scaled coordinates) is an
effective Gouy-Chapman length, characterizing the ionic solution far away
from the plate. 

The asymptotic density profile found in our approximation,
$n(z) = c({\Xi})/(z+1)^2$, 
is different from Eq.~(\ref{nasym}). To understand this difference let us
explain first the asymptotic behavior of $f(z)$:
it decays in the TCMF model as $1/(z+1)$ because 
beyond the correlation hole the test charge is surrounded by
an ion density of the form $n_{\rm PB} = 1/(z+1)^2$. This form
is different from the profile $n(z)$ that is eventually obtained
by integrating $f(z)$ -- 
an inconsistency which is the source of the difference between
$f(z)$ in our approximation and the hypothesized form
$f(z) \sim 1/(z+b)$ (see also the discussion 
in Appendix \ref{ap:contact}). 

The behavior of our approximate $f(z)$ leads to a decay of 
$n(z)$ of the form
$1/(z+1)^2$. The normalization condition for $n(z)$ is 
then enforced through a small prefactor $c({\Xi})$.
In comparison, in the hypothesized form (\ref{nasym}) 
the prefactor must be 1 and the normalization
is achieved by a large value of $b$. 
Note that $b$ must be an extremely large number for large
values of $\Xi$: due to the 
exponential decay of $n(z)$ at $z \lesssim \sqrt{\Xi}$
the logarithm of $b$ must be at least of order $\sqrt{\Xi}$.

Further insight on the behavior at $z > \Xi$ can be obtained using
the exact equation (\ref{dlnrhoex}):
\begin{equation}
\frac{\rm d}{{\rm d}z}{\rm log}\left[n(z)\right] = -f(z) 
\label{exact}
\end{equation}
where $f(z)$ is now the mean (thermally averaged) electrostatic 
force acting on a test charge at distance $z$ from the plate,
in the exact theory. 

For the mean field form $n_{\rm asym}(z)$ to be correct, the contribution
to $f(z)$ coming from the influence of the test charge on its environment
must be small compared to the mean field force, which is given
by $1/(z+b)$. Following our results from the previous section, the
former quantity is given by ${\alpha}\Xi/z^2$, where $\alpha$
is of order unity. 
Using the mean field equation (\ref{PBeq}) 
we obtained $\alpha = 3/4$; 
in the exact theory, and for large $\Xi$, where ions form a much
more localized layer close to the plane than in mean field,
it is plausible that $\alpha =1/4$, as in the
force acting on a test charge next to a conducting 
surface \cite{Shklovskii,MoreiraNetz02}. In any case, for 
Eq.~(\ref{nasym}) 
to represent correctly the decay of $n(z)$ we must have
\begin{equation}
\frac{\alpha\Xi}{z^2} < \frac{1}{z+b}
\label{condition}
\end{equation}
leading to the result $z \gtrsim (\Xi b)^{1/2}$ which is exponentially
large due to $b$. Up to this crossover distance the decay of $n(z)$ 
is dominated by the correlation-induced interaction with the ions
close to the plate.

We conclude that for a very large range
of $z$ values
the decay of $n(z)$ must be different from Eq.~(\ref{nasym}).
At the same time, a mean field argument is probably applicable, 
because the density of ions is very small in this regime: we
may presume that the contribution to $f(z)$ can be divided into two
parts -- one part, coming from ions far away from the plate, which
is hardly influenced by the test charge; and a second part, coming from
ions close to the charged plate, where the
test charge influence on $f(z)$ is given 
by $\alpha \Xi/z^2$. This reasoning leads to the following differential
equation for $n(z)$:
\begin{equation}
\frac{{\rm d}^2}{{\rm d}z^2}{\rm log}[n(z)] = 
2 n(z) + \frac{2\alpha\Xi}{z^3}
\label{newmf}
\end{equation}
whose detailed derivation is given 
in Appendix \ref{ap:derivation2}. By defining 
$n(z) = {\rm exp}(-\phi+\alpha\Xi/z)$ 
Eq.~(\ref{newmf}) is recast in the form:
\begin{equation}
\frac{{\rm d}^2\phi}{{\rm d}z^2} = -2 {\rm exp}\left(
-\phi+\frac{\alpha\Xi}{z}\right) = -2n(z)
\label{newmf1}
\end{equation}
showing that mean field theory is applicable, 
but an external potential
$-\alpha\Xi/z$, coming from the ions close to the plate,
must be included in the PB equation. In practice, for large $\Xi$ this
equation will lead to a decay of the form 
$n(z) \sim {\rm exp}(\alpha \Xi/z)$ as suggested also in 
Refs.~\cite{Shklovskii,MoreiraNetz02}, while
a crossover to an algebraic decay will occur only 
at a distance of at least
$\left[\Xi{\rm exp}(\sqrt{\Xi})\right]^{1/2}$ where 
Eq.~(\ref{condition}) has been used and prefactors
of order unity, inside and outside the exponential, are omitted.
A numerical solution of Eq.~(\ref{newmf1}) may be useful in 
order to describe the ionic layer at intermediate
values of $\Xi$ (of order 10 - 100), where both mean field and 
correlation-induced
forces are of importance at moderate distances from the plate.
In order to test this idea quantitatively more data 
from simulation is required.

Finally we discuss the case where $z >\sqrt{\Xi}$ but
$z$ is not large compared to $\Xi$. Let us also assume that
$\Xi$ is very large, so that most of the ions are much closer to the 
plate than a test charge fixed at $z\hat{\bf z}$. Within the TCMF model
the effect of the test charge on ions close to the plate is nonlinear,
leading to the scaling result of Eq.~(\ref{collapse}). Similarly,
in the exact theory it is not clear whether the correlation-induced
force acting on the test charge is of the form $\alpha\Xi/z^2$, since
the effect of the test charge on ions close to the plate is
not a small perturbation. Hence we believe that
the relevance of Eq.~(\ref{newmf1}) for $z < \Xi$,
and the behavior of $f(z)$ in this regime, 
merit further investigation
\footnote{
In order to improve over TCMF results
it may by useful to treat mobile ions close to the plane 
as a two dimensional layer, while
going beyond the mean field
approximation of the TCMF model in their treatment. The response
to the test charge at ${\bf r} = z_0\hat{\bf z}$
may then help understand the decay of $n(z)$ in the full
three dimensional problem, through Eq.~(\ref{dlnrho1}). 
Such a calculation is beyond the scope of the present work.
}.

%=============================================================
\section{Conclusion}
%=============================================================

In this work we showed how ion correlation effects can be studied
by evaluating the response of the ionic solution to the presence
of a single test particle. Although we calculated this response using
a mean field approximation, we obtained exact results in the limits
of small and large $\Xi$, and qualitative agreement with simulation
at intermediate values. 

The approach taken in this work demonstrates that for highly correlated
ionic liquids it is essential
to treat the particle charge in a non-perturbative manner. Once a 
single ion is singled out, even a mean field approximation applied 
to the other ions provides useful results. This scheme,
called the test-charge mean field (TCMF) model,
provides a relatively
simple approximation, capturing the essential effects of strong
correlations -- to which more sophisticated treatments can be compared.

Technically what is evaluated in this work is the
ion-surface correlation function. 
Consideration of correlation functions
of various orders leads naturally to liquid state theory
approximations \cite{BlumHenderson},
some of which are very successful in describing
ionic liquids \cite{KjelMar}. In particular
these approximations usually treat the ion-ion correlation function
in a more consistent manner than the approximation used in this work,
thus possibly alleviating some of the undesirable features of the
TCMF model presented here, such as the violation
of the contact theorem. The main advantage of the TCMF
model is its simplicity, allowing the
behavior of $n(z)$ to be understood in all the range of coupling 
parameter values in terms of $f(z)$, the force acting on a test charge.
Furthermore, the exact equation (\ref{dlnrhoex}), which does not involve
any approximation, is a useful tool in assessing correlation 
effects -- as was done, for example, in this work in
the end of Sec.~\ref{sec:further}. 

It will be useful to summarize the main results obtained in 
this work. First, the exact equation (\ref{dlnrhoex}) provides a 
simple explanation of the exponential decay of the density profile
in the strong coupling limit. In light of this equation,
exponential decay is expected as
long as the test charge is decoupled from the rest of the ionic
solution. Note that there is no necessity for long
range order to exist in order for the exponential decay to occur,
as was emphasized also in Ref.~\cite{Shklovskii}. 
Indeed, within our TCMF model the ion distribution around
a test charge does not display any long range order (see Fig.~5(a))
and yet simulation results, in particular in the strong coupling limit,
are recovered very successfully. 

Second, the characteristic 
size of the correlation hole around an ion close to
the plane, $\sqrt{\Xi}$, plays an important role in determining
the density profile. For 
very large $\Xi$ the profile decays exponentially
up to $z \lesssim \sqrt{\Xi}$, beyond which a crossover
to a less rapid decay occurs. For intermediate values
of the coupling parameter $z = \sqrt{\Xi}$ is still an approximate 
boundary between regimes of different behavior of $n(z)$, 
but the density profile at $z < \sqrt{\Xi}$ does not decay
in the simple exponential form ${\rm exp}(-z)$. In this sense 
one cannot speak of a region close to the plate where strong
coupling results are valid. 

For $z \gtrsim \Xi$ our approximation predicts a transition
to an algebraic decay of $n(z)$, of the form $c(z)/(z+1)^2$,
where the prefactor $c(z)$ is exponentially small for large 
$\Xi$. A different asymptotic behavior of the form 
$1/(z+b)^2$ is probably valid at very large $z$, but is not
predicted by the TCMF model. Arguments presented
in Sec.~\ref{sec:further}, 
based on the exact equation (\ref{dlnrhoex}), 
lead to the conclusion that for large $\Xi$
the latter form (with a constant value of $b$) 
can only be valid at extremely large values of $z$, while
suggesting that at all distances from the plate larger than
$\Xi$ a modified mean field equation, Eq.~(\ref{newmf1}), is valid.
This equation, matched with the behavior of the ion distribution
close to the charged plate, ultimately determines the value of
the effective Gouy-Chapman length $b$.

Finally, as a by-product of the analysis of 
Sec.~\ref{sec:far}, we obtain
scaling results for the interaction of a high-valent counterion 
with a charged plane immersed in a weakly correlated ionic liquid.
All the results of Sec.~\ref{sec:far} and in particular the scaling 
form (\ref{collapse}), valid for $z \gtrsim \sqrt{\Xi}$, can be regarded
as exact in such a system.

Our approach can be easily generalized to more complicated geometries
than the planar one, although the practical solution of the PB equation
with a test charge may be more difficult in these cases. Other natural
generalizations are to consider non-uniformly charged surfaces
and charged objects in contact with a salt solution. Beyond the TCMF
approximation of Eqs.~(\ref{PBeq}), (\ref{model}), 
and (\ref{modelnorm}), the exact equation 
(\ref{dlnrhoex}) always applies and can be a very useful tool for assessing 
correlation effects near charged macromolecules of various geometries.

We conclude by noting that important questions remain open regarding the
infinite planar double layer, which is the most simple model of 
a charged macroion in solution. One such issue, on which the present 
work sheds light, is the crossover from a strongly coupled liquid
close to the charged plate to a weakly correlated liquid further away.
In particular, the precise functional dependence of the effective 
Gouy-Chapman length $b$ on $\Xi$ is still not known.
More simulation results, in particular at large distances from
the charged plate, and a direct evaluation of $f(z)$ from simulation,
may be useful in order to gain further understanding and to test some
of the ideas presented in this work.
Another important issue, which has not been addressed at all in the
present work, is the possible emergence of a crystalline 
long range order parallel to the plane at sufficiently large
values of the coupling parameter. Although plausible arguments
have been presented for the occurrence of such a phase transition 
at $\Xi \gtrsim 3\times 10^4$ \cite{Shklovskii}, 
its existence has not been proved.

\begin{acknowledgments}
We wish to thank 
Andre Moreira and Roland Netz for providing us with their simulation
data. We thank Boris Shklovskii for helpful comments and
thank Nir Ben-Tal, Daniel Harries, Dalia Fishelov, 
and Emir Haliva for useful advice regarding the numerical 
solution of the partial differential equation. 
Support from the Israel Science Foundation (ISF) under grant no. 210/02
and the US-Israel Binational Foundation (BSF) under grant no. 287/02 is
gratefully acknowledged.
\end{acknowledgments}

\appendix

%=============================================================
\section{Mean field free energy}
%=============================================================
\label{ap:freeenergy}

In this appendix we show how the mean field free energy (\ref{FPB}) is
obtained as an approximation to $F(z_0)$, Eq.~(\ref{Fz0}). 
We start from a general
expression for the grand canonical potential of 
an ionic solution interacting with an external and fixed charge
distribution $\sigma({\bf r})$. In the mean field 
approximation \cite{BurakAndelman00, NetzOrland00},
\begin{eqnarray}
\Omega & = &  \int {\rm d}^3 {\bf r}\, \left\{
-\frac{1}{8\pi l_B q^2}\left[{\bf \nabla}\phi({\bf r})\right]^2
+\frac{\sigma({\bf r})\phi({\bf r})}{q}
\right. \nonumber \\
& & - \left.
\lambda \Theta({\bf r}){\rm e}^{-\phi({\bf r})}
\right\}
\label{OmegaPB}
\end{eqnarray}
where $q$ is the valency of the ions, $\lambda$ is the fugacity,
$\Theta({\bf r})$ is equal to $1$ in the region accessible to the ions and
to zero elsewhere (equal in our case to $\theta(z)$, the Heaviside
function),
and $\Omega$ is given in units of $k_B T$. Requiring an extremum 
with respect to $\varphi$, the reduced electrostatic potential,
yields the PB equation which determines
the electrostatic potential and the actual value of $\Omega$.
We use equation (\ref{OmegaPB}), which is given in
the grand-canonical ensemble, 
because it is widely used in the literature 
\cite{BurakAndelman00, NetzOrland00}. In Ref.~\cite{NetzOrland00}
Eq.~(\ref{OmegaPB}) is derived systematically
as the zero-th order term in a loop expansion
of the exact partition function. 

Inspection of Eq.~(\ref{Fz0}) shows that it describes 
the free energy of an ionic solution interacting with an 
external charge distribution having the following 
parameters,
\begin{equation}
\begin{array}{llll}
\mbox{valency} & 
q & = & \sqrt{\Xi} 
\\
\mbox{Ext. potential \ } & 
\sigma({\bf r}) & = & 
{\displaystyle
\frac{1}{\sqrt{\Xi}}\left[
-\frac{1}{2\pi}\delta(z)+\Xi \delta({\bf r}-z_0{\bf\hat{z})}
\right]}
\\
\mbox{Bjerrum length}
& l_B & = & 1
\end{array}
\end{equation}
In the second line (external potential) the first term comes
from the uniformly charged plate and the second term comes
from the fixed test charge. Plugging these values in 
Eq.~(\ref{OmegaPB}) yields,
\begin{eqnarray}
\Omega & = & \frac{1}{\Xi}
\int{\rm d}^3{\bf r}\left\{
-\frac{1}{8\pi}({\bf\nabla}\varphi)^2
-\lambda\theta(z){\rm e}^{-\varphi}
\right. \nonumber \\ & + & \left.
\varphi\left[-\frac{1}{2\pi}\delta(z)
+\Xi\delta({\bf r}-z_0{\bf z})
\right]
\right\}
\label{omegaions}
\end{eqnarray}
In order to obtain Eq.~(\ref{FPB}) two modifications are required. First, we need
to return to the canonical ensemble by adding $\mu N$ to $\Omega$, where $N$ is the
total number of ions. Noting that $\mu = {\rm log}\lambda$ and that
from charge neutrality
$
q N = -\int{\rm d}^3{\bf r}\sigma({\bf r}),
$
this modification yields the extra term that is proportional to $\log \lambda$
in Eq.~(\ref{FPB}).
Second, we note that $\Omega$ includes the Coulomb self-energy of the charged
plane and of the test charge. This infinite term does not depend on $z_0$ and
should be subtracted from $\Omega$ since it is
not included in the definition of $F(z_0)$, Eq.~(\ref{Fz0}).

We finally note that the results of this Appendix can also be obtained
directly from the canonical partition function, as expressed
by Eq.~(\ref{Fz0}).

%===========================================================
\section{Derivation of Identity (\ref{dlnrho})}
%===========================================================
\label{ap:derivation}

We would like to evaluate the variation
$\delta F_{\rm PB}(z_0)/\delta z_0$, where
$F_{\rm PB}$ is given by Eq.~(\ref{FPB}). 
Note that $\varphi$ itself depends on
$z_0$. However the first derivative of $F_{\rm PB}$ with respect to
$\varphi({\bf r})$ is zero. Hence the only contribution to 
$\delta F_{\rm PB}/\delta z_0$ comes from the explicit dependence
on $z_0$:
\begin{eqnarray}
\frac{\delta F_{\rm PB}[z_0]}{\delta z_0} & = & 
\frac{1}{\Xi}\int {\rm d}^3{\bf r}\,(\varphi-{\rm log}\lambda)
\Xi \frac{\partial}{\partial z_0}\delta({\bf r}-z_0{\bf\hat{z}})
\nonumber \\
& = &  -\left.\frac{\partial\varphi}{\partial z}
\right|_{\displaystyle {\bf r}=z_0\hat{\bf z}}
\end{eqnarray}
It is also instructive to derive this identity within the exact
theory. Equation (\ref{Fz0}) can be written as follows:
\begin{equation}
{\rm exp}[-F(z_0)] = \frac{1}{(N-1)!}
\int \prod_{i=1}^{N-1}{\rm d}^3{\bf r}_i\,
{\rm exp}\left(-H_{z_0}\{{\bf r}_i\}\right)
\end{equation}
where the $N$-th charge is fixed at ${\bf r}=z_0\hat{\bf z}$:
\begin{equation}
H_{z_0}\{{\bf r}_i\} = 
-z_0 - \sum_{i=1}^{N-1}z_i - \sum_{i=1}^{N-1}
\frac{\Xi}{\left|{\bf r}_i-z_0\hat{\bf z}\right|}
- \sum_{j>i}
\frac{\Xi}{\left|{\bf r}_i-{\bf r}_j\right|}
\end{equation}
Differentiating with respect to $z_0$ yields:
\begin{equation}
\frac{\delta F(z_0)}{\delta z_0} = 
-\left< \frac{\partial H_{z_0}}{\partial z_0}\right>
= -\left<
-1+\sum_{i=1}^{N-1}\frac{\Xi(z_0-z_i)}
{\left|{\bf r}_i-z_0{\bf\hat{z}}\right|^3}\right>
\end{equation}
where the averaging is performed over all configurations of the $N-1$ ions
with the weight ${\rm exp}(-H_{z_0}\{{\bf r}_i\})$. This quantity is the 
mean electrostatic field acting on a test charge at 
$z_0 {\bf \hat{z}}$.

%============================================================
\section{Numerical scheme}
%============================================================
\label{ap:numerical}

Numerically solving a non-linear PDE such as (\ref{PBeq}) 
requires careful examination of the solution behavior.
The purpose of this section is to explain the numerical
scheme used in this work, and in particular the parameters required
to obtain a reliable solution.

\subsubsection*{Finite cell}

\begin{figure}
\scalebox{0.45}{\includegraphics{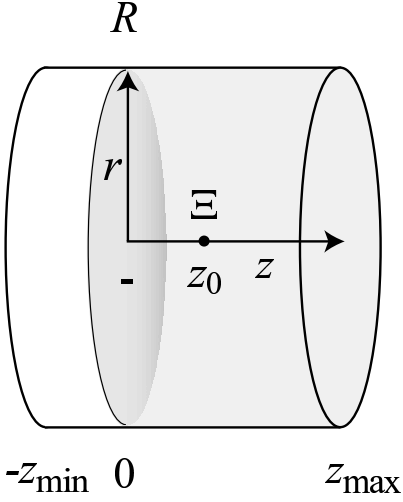}}
\caption{
Schematic illustration of the setup used to solve
numerically the PDE (\ref{PBeq}). The equation is solved in a finite cylindrical 
cell, extending from $-z_{\rm min}$ to $z_{\rm max}$ in the $z$ axis and
from $0$ to $R$ in the $r$ axis. The charged plane is at $z = 0$ and ions
are only present for $z > 0$. Neumann boundary conditions 
$({{\bf \nabla}\varphi}\cdot\hat{\bf n} = 0)$ are imposed at the cell boundaries.
The test charge is at ${\bf r} = z_0\hat{\bf z}$ 
}
\end{figure}
We solve Eq.~(\ref{PBeq})
as a two dimensional problem in the coordinates $r$ and
$z$, making use of the symmetry of rotation around the $z$ axis. 
The problem is defined within a finite cell
of cylindrical shape, shown schematically in
Fig.~11. 
The negatively charged plate is at $z = 0$ and ions are only allowed
in the region $z > 0$, marked as the gray-shaded region in the plot.

We impose a boundary condition of zero electrostatic field,
\begin{equation}{\rm \nabla}\varphi\cdot \hat{\bf n} = 0,
\label{boundary}
\end{equation}
at the cell boundaries: $z = -z_{\rm min}$,  $z = z_{\rm max}$, and
$r = R$. The cell size, as determined by these boundaries, 
must be sufficiently large, as will be further discussed below.

In the numerical solution it is necessary to solve 
$\varphi$ for the 
electrostatic potential at positive as well as negative $z$
\footnote{This situation is different from that of the
PB equation with no test charge, in which the electrostatic
potential depends only on $z$. The boundary condition
at $z = 0$, for the case of zero dielectric
constant at $z < 0$, is then sufficient in order to solve
for the potential at $z > 0$.}.
Note that a boundary condition such as (\ref{boundary})
at $z = 0$ would correspond to zero dielectric constant at $z < 0$,
while we are interested in the case of continuous 
dielectric constant across the plate.

\subsubsection*{Differential equation}

The source term in Eq.~(\ref{PBeq}) diverges at $z = 0$ and 
at ${\bf r} = z_0\hat{\bf z}$. We avoid this difficulty by 
shifting the potential:
\begin{equation}
\varphi = \psi + |z| + \frac{\Xi}{|{\bf r}-z_0\hat{\bf z}|}
\label{psidef}
\end{equation}
and solving for $\psi$, which is the potential due only to the mobile ions. 
The equation for $\psi$, 
\begin{equation}
{\bf\nabla}^2\psi = 
-4\pi \lambda\theta(z){\rm exp}
\left(-\psi-z-\frac{\Xi}{|{\bf r}-z_0\hat{\bf z}|}\right)
\label{PBeqpsi}
\end{equation}
is solved with a Neumann boundary condition for $\psi$, 
derived from
Eqs.~(\ref{boundary}) and (\ref{psidef}). Note that, 
unlike $\varphi$, $\psi$ is well behaved at $z_0\hat{\bf z}$.
The nonlinear equation (\ref{PBeqpsi}) can be solved by iterative solution of
a linear equation (see, for example, \cite{Harries,CarnieChan94}),
\begin{eqnarray}
{\bf\nabla}^2\psi_n
\nonumber & = & -4\pi \lambda\theta(z){\rm exp}
\left(-\psi_{n-1}-z-\frac{\Xi}{|{\bf r}-z_0\hat{\bf z}|}\right)
\nonumber \\
& \times & \left[1-\left(\psi_n-\psi_{n-1}\right)\right]
\label{PBeqit}
\end{eqnarray}
where $\psi_n$ represents the $n$-th iteration.

\subsubsection*{Grid and solution method}

In the coordinates $r$,$z$ the cylindrical 
cell is a rectangular domain, 
\begin{displaymath}
[0,R]\times [-z_{\rm min},z_{\rm max}].
\end{displaymath}
We use bi-cubic Hermite collocation \cite{Houstis} in this domain in order
to translate the PDE into a set of linear algebraic equations
on a grid. 
These equations are
then solved using Gauss elimination with scaled partial pivoting. 
Storing the band matrix representing the linear equations requires 
approximately $48 \times N_r^2\times N_z$ 
memory cells, where $N_r$ and $N_z$ are the number
of grid points in the $r$ and $z$ directions, respectively
\cite{Houstis}. Because this number can be very large, 
it is essential to use a variably spaced grid in both
of the coordinates. We use the following
scheme:

\textbf{r coordinate:} In the absence of a test charge, an arbitrarily 
coarse grid can be used in the $r$ direction, due to the 
translational symmetry parallel to the charged plane. In our case
(where a test charge is present) the
grid spacing is determined by the distance from the test charge, as
follows,
\begin{equation}
\frac{{\rm d}n}{{\rm d}r} = \frac{n_r}{r+r_{\rm grid}},
\label{gridr1}
\end{equation}
where $n_r$ and $r_{\rm grid}$ are two fixed parameters, while
$n$ stands for the grid point index and ${\rm d}n/{\rm d}r$
is the number of grid points per unit increment of the radial 
coordinate. This spacing 
is approximately uniform up to the threshold 
$r_{\rm grid}$, whereas for $r \gg r_{\rm grid}$ it is 
proportional to $1/r$. The grid points are then of the form
\begin{equation}
r_n = r_{\rm grid}\times \left[{\rm exp}\left(\frac{n}{n_r}\right)-1
\right]
\label{gridr2}
\end{equation}
In practice $r_{\rm grid}$ is chosen approximately
proportional to $\sqrt{\Xi}$, in order to allow the structure of the
correlation hole to be represented faithfully.

{\bf z coordinate:} In this coordinate the grid spacing is influenced
by the distance from the charged plate as well as the distance
from the test charge. We describe separately the spacing determined
from each of these two criteria; the actual grid is obtained by using
the smaller of the two spacings at each point.

(\textit{i}) Distance from the plate: we use a
grid spacing proportional to the derivative of
$\varphi_{\rm PB}(z)$:
\begin{equation}
\frac{{\rm d}n}{{\rm d}z} \propto \frac{2}{z+1}
\label{gridz}
\end{equation}
Ignoring, for the moment, the distance from the test charge, 
Eq.~(\ref{gridz}) leads to grid points of the form
\begin{equation}
z_{n} = {\rm exp}(n \cdot D)-1
\end{equation}
where the parameter $D$ is the grid spacing close to the charged plate.
A similar scheme is used in the negative $z$ half-space.

(\textit{ii}) Distance from the test charge: we use a form similar to 
(\ref{gridr1}),
\begin{equation}
\frac{{\rm d}n}{{\rm d}z} = \frac{n_z}{|z-z_0|+z_{\rm grid}},
\end{equation}
In practice, the threshold $z_{\rm grid}$ is chosen
proportional to $\Xi$, in contrast to $r_{\rm grid}$ which
is chosen proportional to $\sqrt{\Xi}$.

\subsubsection*{Parameters}

The parameters that were used to obtain the numerical
results presented in this
work are summarized in Table~2. We compared our results with 
those obtained with (a) Increasing $z_{\rm min}$, $z_{\rm max}$, and $R$ 
by a factor of 10; and (b) decreasing the grid spacing by a 
factor of 2, both in
the $r$ and in the $z$ coordinates. The influence of these changes was
found to be negligible on all the data presented in this work.

\begin{table}
\caption{Parameters used in numerical solution of the PDE.}
\begin{ruledtabular}
\begin{tabular}{lllllllll}
$\Xi$ & $z_{\rm min}$ & $z_{\rm max}$ & $R$ & $D$ & $z_{\rm grid}$ &
$n_z$ & $r_{\rm grid}$ & $n_r$ \\
\hline
$10^4$ & $10^4$ & $10^5$ & $10^5$ & 0.2 & 5000 & 5 & 100 & 5 \\
$10^3$ & $10^4$ & $10^5$ & $10^5$ & 0.2 & 500 & 4 & 33 & 4 \\
$10^2$ & $10^3$ & $10^4$ & $10^4$ & 0.2 & 50 & 4 & 10 & 4 \\
10 & $10^3$ & $10^4$ & $10^4$ & 0.2 & 5 & 4 & 3 & 4 \\
1 & $4\times 10^2$ & $4 \times 10^3$& $4 \times 10^3$ & 0.2 & 
0.5 & 4 & 1 & 4 \\
0.1 & 80 & 800 & 800 & 0.2 & 0.05 & 4 & 0.2 & 4 \\
\end{tabular}
\end{ruledtabular}
\end{table}

%============================================================
\section{Contact theorem}
%============================================================
\label{ap:contact}

In this appendix we derive the contact theorem \cite{CarnieChan81} 
in a way that highlights the reason why it is not obeyed in
our approximation. We start from an exact expression for the free
energy,
\begin{equation}
F = -{\rm log}\int_a^{\infty}{\rm d}z'\,
{\rm exp}[-F(z')]
\end{equation}
where the charged plate is at $z = a$. This plate position 
can be chosen arbitrarily, hence $\partial F/\partial a = 0$:
\begin{equation}
0 = \left. \frac{\partial F}{\partial a} \right|_{a = 0}= 
n(0) - \int_a^{\infty}{\rm d}z'\,n(z')
\frac{\partial F}{\partial z}(z')
\end{equation}
where we used the relations
\begin{equation}
n(z) = \frac{{\rm exp}[-F(z)]}{\int_0^{\infty}{\rm d}z'\,
{\rm exp}[-F(z')]}
\end{equation}
and 
\begin{equation}
\left.\frac{\partial F(z)}{\partial a}\right|_{\displaystyle z} = 
- \left.\frac{\partial F(z)}{\partial z}\right|_{\displaystyle a}
\end{equation}
We now use Eq.~(\ref{dlnrhoex}) to obtain,
\begin{equation}
n(0) - \int_0^{\infty}n(z)
\left.\frac{\partial}{\partial z}
\left<\varphi({\bf r}; z)\right>
\right|_{\displaystyle {\bf r} = z_0 \hat{\bf z}} = 0
\end{equation}
The second term in this equation is the average electrostatic
force acting on the ions per unit area. This force can be separated into
two contributions. The first one, exerted by the charged plane, is 
equal to $-\int{\rm d}z'\, n(z') = -1$ because the plane applies an 
electrostatic force which does not depend on $z'$ and
is equal to unity in our rescaled coordinates.
The remaining contribution to the force, 
exerted by the ions on themselves, must be
zero due to Newton's third law, leading to the result $n(0) = 1$.

The discussion up to now was exact. It also applies to PB theory,
where the self consistency of the mean field approximation
ensures that Newton's third law is obeyed.
On the other hand, within our approximation
the force exerted by the ions on themselves,
\begin{displaymath}
\int_0^{\infty}n(z) \left[f(z) - 1 \right]
\end{displaymath}
is not zero. This inconsistency can be traced to a more fundamental 
inconsistency which is briefly described below.

The probability to find the test
charge at ${\bf r} = z\hat{\bf z}$ and a mobile ion 
at ${\bf r'}$ is proportional, in our approximation, to
\begin{equation}
n(z) n(z') g({\bf r},{\bf r'}) \equiv
n(z){\rm exp}\left[-\varphi({\bf r'};z)\right] 
\end{equation}
In the exact theory the probability
to find two ions at ${\bf r}$ and ${\bf r'}$ must be symmetric
with respect to exchange of ${\bf r}$ and ${\bf r'}$. On the
other hand the correlation function
$g({\bf r},{\bf r'})$, as defined above,
is not symmetric. In other words, the ion-ion correlation 
function in the TCMF model is not symmetric.

%============================================================
\section{Small $\Xi$ expansion}
%============================================================
\label{ap:smallxi}

The recovery of mean field results at small $\Xi$ was 
demonstrated and explained in Sec.~\ref{sec:model}. Here we derive this
result formally as an expansion in powers of $\Xi$. The advantage
of this formal expansion is that it
allows us to find also the first order correction to the PB
profile within our model.

We expand $\Xi F_{\rm PB}$, Eq.~(\ref{FPB}), up to second order in $\Xi$:
\begin{equation}
\Xi F_{\rm PB}(z_0) = F_0 + \Xi F_1(z_0) + \Xi^2 F_2(z_0) + \cdots
\end{equation}
The zero-th order term, $F_0$, does not depend on $z_0$ and is the
PB free energy of a charged plane in contact with its counterions, 
without a test charge. In order to evaluate the following terms,
we also expand $\varphi$ in powers of $\Xi$:
\begin{equation}
\varphi({\bf r};z_0) = \varphi_0({\bf r})+\Xi\varphi_1({\bf r};z_0)+
\Xi^2\varphi_2({\bf r};z_0)+\cdots
\end{equation}
To zero-th order we have from Eqs.~(\ref{FPB}) and (\ref{PBeq}):
\begin{equation}
F_0 =
\int{\rm d}^3{\bf r}\left\{
-\frac{1}{8\pi}({\bf\nabla}\varphi_0)^2
-\lambda\theta(z){\rm e}^{-\varphi_0(z)}
\right\}
\label{f0}
\end{equation}
where
\begin{equation}
{\bf\nabla}^2\varphi_0 = \frac{{\rm d}^2\varphi_0}{{\rm d}z^2}=
-4\pi \lambda\theta(z){\rm e}^{-\varphi_0} 
\label{eq0}
\end{equation}
is the potential due to counterions in the PB approximation.
The first order term in the free energy, $F_1$, is found by expanding
equation (\ref{FPB}) in $\Xi$. This expansion includes two contributions,
the first from $\varphi_1$ and the second from the explicit dependence
on $\Xi$ in Eq.~(\ref{FPB}).
The first contribution vanishes because $\varphi_0$ is an extremum
of the zero-th order free energy, leaving only the second contribution:
\begin{eqnarray}
F_1(z_0) & = &\int{\rm d}^3{\bf r}\,
\varphi_0({\bf r})\delta({\bf r}-z_0\hat{\bf z})
= \varphi_0(z_0)
\end{eqnarray}
Returning to our approximation for $n(z)$, given by Eq.~(\ref{model}),
we find that:
\begin{eqnarray}
n(z) & = & \frac{1}{Z}{\rm exp}\left[-F_{\rm PB}(z)\right]
= \frac{1}{Z}{\rm exp}\left[-\frac{F_0}{\Xi} - \varphi_0(z)\right] 
\nonumber \\
& = & 
\frac{1}{Z_0}{\rm exp}\left[-\varphi_0(z)\right],
\end{eqnarray}
where $Z_0$ is found from the normalization condition (\ref{modelnorm}).
To leading order in $\Xi$, $n(z)$ is equal to the PB density profile,
as expected:
\begin{equation}
n(z)=
n_{\rm PB}(z) = \frac{1}{Z_0}\,{\rm exp}[-\varphi_0(z)] 
= \frac{1}{(z+1)^2}
\end{equation} 
where $Z_0$ is a normalization constant. 
The next order term in the expansion of $f$ can be found on similar
lines as $F_1(z)$, and is equal to
\begin{equation}
F_2(z_0) = \frac{1}{2}
\delta\varphi_1(z_0\hat{\bf z}; z_0)
\label{f2}
\end{equation}
where $\delta\varphi$ is the difference between the first order
correction to $\varphi$ and the bare potential of the test charge:
\begin{equation}
\delta\varphi_1({\bf r}) = \varphi_1({\bf r}) 
- \frac{1}{|{\bf r}-z_0\hat{\bf z}|}
\end{equation}
The first order term in the expansion of $\varphi$, 
$\varphi_1({\bf r}; z_0)$ 
is the solution of the differential equation:
\begin{equation}
\left[{\nabla}^2-4\pi\lambda{\rm e}^{-\varphi_0}\right]
\varphi_1 = -4\pi\delta({\bf r}-z_0\hat{\bf z})
\label{varphi1}
\end{equation}
The function $\delta\varphi_1({\bf r})$ 
arises also in the systematic loop expansion 
of the free energy around the mean
field solution \cite{NetzOrland00}. Its value at ${\bf r} = z_0\hat{\bf z}$
is given by \cite{NetzOrland00}:
\begin{eqnarray}
& & \delta\varphi_1(z_0\hat{\bf z}; z_0) \equiv g(z_0)
 = \frac{1}{2(z_0+1)^2}
\times \nonumber \\ & &
\left\{ 
i{\rm e}^{(1-i)z_0}E_1\left[(1-i)z_0\right](1+iz_0)^2
\right.
\nonumber \\ 
& & \left.
-i{\rm e}^{(1+i)z_0}E_1\left[(1+i)z_0\right](1-iz_0)^2
-4z_0 \right\}
\label{gofz}
\end{eqnarray}
where $E_1[x]$ is the exponential-integral function \cite{AStegun}.
Using Eqs.~(\ref{model}) and (\ref{f2})
we find that up to first order in $\Xi$ the density profile
is given by:
\begin{equation}
n(z) = n_{\rm PB}(z) + \Xi n_1(z)
\end{equation}
where
\begin{equation}
n_1(z) = \left[N_1 - \frac{1}{2}g(z).
\right]n_{\rm PB}(z)
\label{n1}
\end{equation}
In this expression $g(z)$ is given by Eq.~(\ref{gofz})
and $N_1$ is obtained from the normalization condition 
(\ref{modelnorm}):
\begin{equation}
N_1 = \frac{1}{2}\int{\rm d}z\,
n_{\rm PB}(z)g(z) \simeq -0.3104
\end{equation}
Note that this is different from the exact expression for the 
first order correction in $\Xi$
\footnote{See Eq.~(64) and Fig.~4 in Ref.~\cite{NetzOrland00}.}, 
which is obtained in the loop expansion and is not reproduced here,
but is shown 
in Fig.~12. 

Figure 12 shows $n_1(z)$ as defined by Eq.~(\ref{n1}) (solid line). The 
symbols show the correction to $n_{\rm PB}$ calculated numerically from TCMF
for $\Xi = 0.1$ and scaled by $1/\Xi = 10$. At this small value of $\Xi$
the linearization provides a very good approximation for the correction
to $n_{\rm PB}(z)$.
\begin{figure}
\scalebox{0.45}{\includegraphics{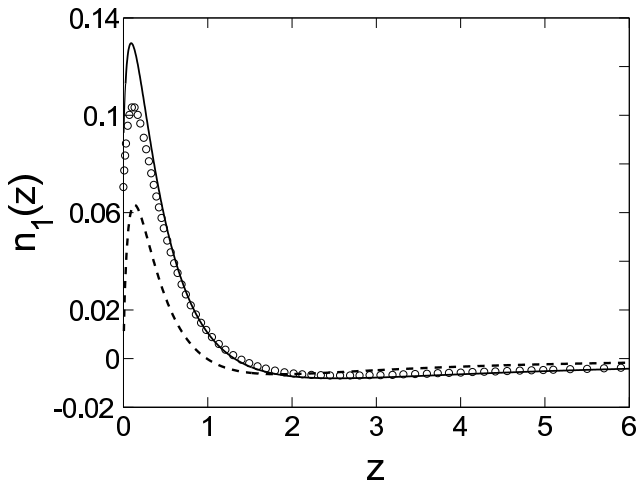}}
\caption{
First order (linear) correction in $\Xi$ to $n(z)$,
as obtained from the test charge mean field approximation, Eq.~(\ref{n1}) (solid line),
compared to the exact first order correction calculated using a loop expansion
\cite{NetzOrland00} (dashed line). The symbols show 
$[n(z)-n_{\rm PB}(z)]/\Xi$ calculated numerically in the test charge mean field
approximation with $\Xi = 0.1$.
}
\end{figure}

The dashed line shows the exact first order correction in $\Xi$ to 
the ion density, obtained from the loop expansion. Comparison 
of the solid and dashed lines shows that the TCMF model does
not capture correctly the exact first order correction. In particular, 
$n_1(0)$ is different from zero in our approximation; in the exact 
correction $n_1(0)=0$ as it must be due to the contact theorem.
It is important to realize that although the exact first order
correction is useful for values of $\Xi$ of order unity and
smaller, the TCMF has a much wider range of validity for 
$\Xi \gtrsim 1$.

\subsubsection*{Proof of Equation (\ref{f1})}

Our purpose here is to prove the first equality of Eq.~(\ref{f1}),
\begin{equation}
f_1(z) = \frac{1}{2} \frac{{\rm d}g(z)}{{\rm d}z}
\end{equation}
where the electrostatic field acting on a test charge is
$-(f_{\rm PB}(z)+\Xi f_1(z)+\ldots)$, \textit{i.e.}, $f_1(z)$ is
the first order term in $\Xi$.
In order to do this, let us consider the correction to the
mean field potential due to an infinitesimal point charge
of magnitude $\Xi$ that is placed at ${\bf r} = z\hat{\bf z}$.
We designate this correction, evaluated at the 
point ${\bf r'}$, as $G({\bf r}, {\bf r'})$. This Green's function is found
by solving Eq.~(\ref{varphi1}) which reads, with a slight change of 
notation:
\begin{equation}
\left[\nabla^2_{\bf r'}
-4\pi\lambda{\rm e}^{-\varphi_0({\bf r'})}\right]
G({\bf r}, {\bf r'}) = -4\pi\delta({\bf r}-{\bf r'})
\label{Geq}
\end{equation}
The electrostatic field acting on the test charge is then
$-f_{\rm PB}(z) - \Xi f_1(z)$, where
\begin{equation}
f_1(z) = \left.\frac{\partial}{\partial z'}G({\bf r},{\bf r'})
\right|_{\displaystyle {\bf r'} = {\bf r}} = \frac{1}{2}
\frac{\partial}{\partial z}G({\bf r},{\bf r}) = 
\frac{1}{2}\frac{{\rm d} g(z)}{{\rm d} z}
\end{equation}
and $g(z)$ is defined in Eq.~(\ref{gofz}).
In the second step we used the symmetry of $G({\bf r},{\bf r}')$
to exchange of ${\bf r}$ and ${\bf r'}$, which follows from
the fact that the operator acting on $G({\bf r},{\bf r}')$
in Eq.~(\ref{Geq}), as well as the right hand side of that
equation, are symmetric with respect to exchange of
${\bf r}$ and ${\bf r'}$.

%============================================================
\section{Mean field equation at large $z$}
%============================================================

\label{ap:derivation2}
We start from the exact identity (\ref{exact}) and would like to evaluate
$f(z)$ for a test particle placed 
at sufficiently large $z$, assuming
also that $\Xi$ is large. The mean field 
electrostatic force acting on the particle is given by
\begin{equation}
f_{\rm MF}(z) = 1 - \int_0^{z}{\rm d}z'\,n(z') + 
\int_z^{\infty}{\rm d}z'\,n(z')
\label{fmf}
\end{equation}
where the first term on the right hand side is the contribution
of the charged plane, the second term is the contribution
of ions between the plane and the test particle, and the third term
is the contribution of the other ions. 
Eq.~(\ref{fmf}) would describe the exact force 
acting on the test particle had it not had
any effect on the distribution of the other ions
in the system. We need to add to this force the contribution
due to the influence of the test charge on the other ions.

Due to the exponential decay close to the plate
the ion layer further than $z = \sqrt{\Xi}$ 
is very dilute. Hence it makes sense
to include in the correlation-induced force only a
contribution
from the ions close to the plate. Estimating this contribution
as $\alpha \Xi/z^2$ we conclude that
\begin{equation}
\frac{{\rm d}{\rm log}n(z)}{{\rm d}z} = -f(z) = 
-f_{\rm MF}(z) - \frac{\alpha \Xi}{z^2}
\end{equation}
Differentiation of this equation 
with respect to $z$ yields Eq.~(\ref{newmf}):
\begin{equation}
\frac{{\rm d}^2{\rm log}n(z)}{{\rm d}z^2} = 
2 n(z) + \frac{2\alpha \Xi}{z^3}
\end{equation}
%

%=======================================================

\end{document}